\documentclass[useAMS,usenatbib,letterpaper]{mn2e}
\usepackage{graphicx}
\usepackage{natbib}
\usepackage{amsmath}
\usepackage{upgreek}
\usepackage{booktabs}

\newcommand\psr{PSR~B0823+26}
\title[LOFAR discovery of a quiet emission mode in \psr]
{LOFAR discovery of a quiet emission mode in \psr}

\author[C.~Sobey, et al.]{
C.~Sobey$^{1,2}$\thanks{E-mail: sobey@astron.nl}, 
N.~J.~Young$^{3,4}$,
J.~W.~T.~Hessels$^{2,5}$,
P.~Weltevrede$^{4}$,
A.~Noutsos$^{1}$,\newauthor
B.~W.~Stappers$^{4}$,
M.~Kramer$^{1,4}$,
C.~Bassa$^{2}$,
A.~G.~Lyne$^{4}$,
V.~I.~Kondratiev$^{2,6}$,\newauthor
T.~E.~Hassall$^{7}$,
E.~F.~Keane$^{8,9}$,
A.~V.~Bilous$^{10}$,
R.~P.~Breton$^{7}$,
J.-M.~Grie\ss{}meier$^{11,12}$,\newauthor
A.~Karastergiou$^{13}$,
M.~Pilia$^{2}$,
M.~Serylak$^{14}$,
S.~ter~Veen$^{10}$,
J.~van Leeuwen$^{2,5}$,
A.~Alexov$^{15}$,\newauthor
J.~Anderson$^{16}$,
A.~Asgekar$^{2,17}$,
I.~M.~Avruch$^{18,19}$,
M.~E.~Bell$^{20}$,
M.~J.~Bentum$^{2,21}$,\newauthor
G.~Bernardi$^{22}$,
P.~Best$^{23}$,
L.~B\^{i}rzan$^{24}$,
A.~Bonafede$^{25}$,
F.~Breitling$^{26}$,
J.~Broderick$^{7}$,\newauthor
M.~Br\"{u}ggen$^{25}$,
A.~Corstanje$^{10}$,
D.~Carbone$^{5}$,
E.~de Geus$^{2,27}$,
M.~de Vos$^{2}$,
A.~van Duin$^{2}$,\newauthor
S.~Duscha$^{2}$,
J.~Eisl\"{o}ffel$^{28}$,
H.~Falcke$^{10,2}$,
R.~A.~Fallows$^{2}$,
R.~Fender$^{13}$,
C.~Ferrari$^{29}$,\newauthor
W.~Frieswijk$^{2}$,
M.~A.~Garrett$^{2,24}$,
A.~W.~Gunst$^{2}$,
J.~P.~Hamaker$^{2}$,
G.~Heald$^{2,19}$,\newauthor
M.~Hoeft$^{28}$,
J.~H\"{o}randel$^{10}$,
E.~J\"{u}tte$^{30}$,
G.~Kuper$^{2}$,
P.~Maat$^{2}$,
G.~Mann$^{26}$,
S.~Markoff$^{5}$,\newauthor
R. McFadden$^{2}$,
D.~McKay-Bukowski$^{31,32}$,
J.~P.~McKean$^{2,19}$,
D.~D.~Mulcahy$^{7}$,
H.~Munk$^{2}$,\newauthor
A.~Nelles$^{10}$,
M.~J.~Norden$^{2}$,
E.~Orr\`{u}$^{2}$,
H.~Paas$^{33}$,
M.~Pandey-Pommier$^{34}$,
V.~N.~Pandey$^{2}$,\newauthor
G.~Pietka$^{13}$,
R.~Pizzo$^{2}$,
A.~G.~Polatidis$^{2}$,
D.~Rafferty$^{24}$,
A.~Renting$^{2}$,
H.~R\"{o}ttgering$^{24}$,\newauthor
A.~Rowlinson$^{20}$,
A.~M.~M.~Scaife$^{7}$,
D.~Schwarz$^{35}$,
J.~Sluman$^{2}$,
O.~Smirnov$^{36,37}$,\newauthor
M.~Steinmetz$^{26}$,
A.~Stewart$^{13}$,
J.~Swinbank$^{5}$,
M.~Tagger$^{11}$,
Y.~Tang$^{2}$,
C.~Tasse$^{38}$,\newauthor
S.~Thoudam$^{10}$,
C.~Toribio$^{2}$,
R.~Vermeulen$^{2}$,
C.~Vocks$^{26}$,
R.~J.~van Weeren$^{22}$,\newauthor
R.~A.~M.~J.~Wijers$^{5}$,
M.~W.~Wise$^{2,5}$,
O.~Wucknitz$^{1}$,
S.~Yatawatta$^{2}$,
P.~Zarka$^{38}$\\ 
$^{1}$Max-Planck-Institut f\"{u}r Radioastronomie, Auf dem H\"{u}gel 69, 53121 Bonn, Germany\\
$^{2}$ASTRON, the Netherlands Institute for Radio Astronomy, Postbus 2, 7990 AA Dwingeloo, The Netherlands\\
$^{3}$School of Physics, University of the Witwatersrand, PO Box Wits, Johannesburg, 2050, South Africa\\
$^{4}$Jodrell Bank Centre for Astrophysics, School of Physics and Astronomy, The University of Manchester, Manchester M13 9PL, UK\\
$^{5}$Anton Pannekoek Institute for Astronomy, University of Amsterdam, Science Park 904, 1098 XH Amsterdam, The Netherlands\\
$^{6}$Astro Space Centre, Lebedev Physical Institute, Russian Academy of Sciences, Profsoyuznaya Str. 84/32, Moscow 117997, Russia\\
$^{7}$School of Physics and Astronomy, University of Southampton, SO17 1BJ, UK\\
$^{8}$Centre for Astrophysics and Supercomputing, Swinburne University of Technology, Mail H30, PO Box 218, VIC 3122, Australia\\
$^{9}$ARC Centre of Excellence for All-sky Astrophysics (CAASTRO), 44 Rosehill Street, Redfern, NSW 2016, Australia\\
$^{10}$Department of Astrophysics/IMAPP, Radboud University Nijmegen, PO Box 9010, 6500 GL Nijmegen, The Netherlands\\
$^{11}$LPC2E - Universit\'{e} d'Orl\'{e}ans/CNRS\\
$^{12}$Station de Radioastronomie de Nan\c{c}ay, Observatoire de Paris - CNRS/INSU, USR 704 - Univ. Orleans, OSUC, 18330 Nan\c{c}ay, France\\
$^{13}$Astrophysics, University of Oxford, Denys Wilkinson Building, Keble Road, Oxford, OX1 3RH, UK\\
$^{14}$Department of Physics \& Astronomy, University of the Western Cape, Private Bag X17, Bellville 7535, South Africa\\
$^{15}$Space Telescope Science Institute, 3700 San Martin Drive, Baltimore, MD 21218, USA\\
$^{16}$Helmholtz-Zentrum Potsdam, GFZ, Department 1: Geodesy and Remote Sensing, Telegrafenberg, A17, 14473 Potsdam, Germany\\
$^{17}$Shell Technology Center, Bangalore, India\\
$^{18}$SRON Netherlands Institute for Space Research, PO Box 800, 9700 AV Groningen, The Netherlands\\
$^{19}$Kapteyn Astronomical Institute, PO Box 800, 9700 AV Groningen, The Netherlands\\
$^{20}$CSIRO Australia Telescope National Facility, PO Box 76, Epping NSW 1710, Australia\\
$^{21}$University of Twente, PO Box 217, 7500 AE Enschede, The Netherlands\\
$^{22}$Harvard-Smithsonian Center for Astrophysics, 60 Garden Street, Cambridge, MA 02138, USA\\
$^{23}$Institute for Astronomy, University of Edinburgh, Royal Observatory of Edinburgh, Blackford Hill, Edinburgh EH9 3HJ, UK\\
$^{24}$Leiden Observatory, Leiden University, PO Box 9513, 2300 RA Leiden, The Netherlands\\
$^{25}$University of Hamburg, Gojenbergsweg 112, 21029 Hamburg, Germany\\
$^{26}$Leibniz-Institut f\"{u}r Astrophysik Potsdam (AIP), An der Sternwarte 16, 14482 Potsdam, Germany\\
$^{27}$SmarterVision BV, Oostersingel 5, 9401 JX Assen, The Netherlands\\
$^{28}$Th\"{u}ringer Landessternwarte, Sternwarte 5, D-07778 Tautenburg, Germany\\
$^{29}$Laboratoire Lagrange, UMR7293, Universit\`{e} de Nice Sophia-Antipolis, CNRS, Observatoire de la C\'{o}te d'Azur, 06300 Nice, France\\
$^{30}$Astronomisches Institut der Ruhr-Universit\"{a}t Bochum, Universitaetsstrasse 150, 44780 Bochum, Germany\\
$^{31}$Sodankyl\"{a} Geophysical Observatory, University of Oulu, T\"{a}htel\"{a}ntie 62, 99600 Sodankyl\"{a}, Finland\\
$^{32}$STFC Rutherford Appleton Laboratory,  Harwell Science and Innovation Campus,  Didcot  OX11 0QX, UK\\
$^{33}$Center for Information Technology (CIT), University of Groningen, The Netherlands\\
$^{34}$Centre de Recherche Astrophysique de Lyon, Observatoire de Lyon, 9 av Charles Andr\'{e}, 69561 Saint Genis Laval Cedex, France\\
$^{35}$Fakult\"{a}t fŸr Physik, Universit\"{a}t Bielefeld, Postfach 100131, D-33501, Bielefeld, Germany\\
$^{36}$Department of Physics and Electronics, Rhodes University, PO Box 94, Grahamstown 6140, South Africa\\
$^{37}$SKA South Africa, 3rd Floor, The Park, Park Road, Pinelands, 7405, South Africa\\
$^{38}$LESIA, UMR CNRS 8109, Observatoire de Paris, 92195 Meudon, France}

\begin{document}

\date{Accepted 2015 May 11. Received 2015 May 7; in original form 2014 September 7}

\pagerange{\pageref{firstpage}--\pageref{lastpage}} \pubyear{2014}

\maketitle

\label{firstpage}

\begin{abstract}

\psr, a 0.53-s radio pulsar, displays a host of emission phenomena
over timescales of seconds to (at least) hours, including nulling,
subpulse drifting, and mode-changing.  Studying pulsars like {\psr}
provides further insight into the relationship between these various
emission phenomena and what they might teach us about pulsar
magnetospheres.  Here we report on the LOFAR discovery that {\psr} has a
weak and sporadically emitting `quiet' (Q) emission mode that is 
over 100 times weaker (on average) and has a nulling fraction forty-times
greater than that of the more regularly-emitting `bright' (B) mode.
Previously, the pulsar has been undetected in the Q-mode, and was
assumed to be nulling continuously.  {\psr} shows a further decrease in average flux just
before the transition into the B-mode, and perhaps truly turns off
completely at these times.  Furthermore, simultaneous observations
taken with the LOFAR, Westerbork, Lovell, and Effelsberg telescopes
between 110\,MHz and 2.7\,GHz demonstrate that the transition between
the Q-mode and B-mode occurs within one single
rotation of the neutron star, and that it is concurrent across the range of
frequencies observed.

\end{abstract}


\section{Introduction}

The average pulse profile shapes of radio pulsars are remarkably
stable when summed over at least a few hundred rotations.  This
reflects their clock-like rotational predictability
\citep[e.g.][]{2012MNRAS.420..361L} and enables their use as precision
physical probes \citep[e.g.][]{2005ASPC..328...25W,2010CQGra..27h4013H,2011ApJ...728...97V}.  
Nonetheless, pulsar emission variability has
been recognised on almost all timescales that are observationally
accessible \cite[e.g.][]{2013IAUS..291..295K}: from nanosecond
`shots', expected to be the quanta of pulsar emission
\citep{2003Natur.422..141H}, to multidecadal variations presumably due
to evolution of the magnetic field \citep{2013Sci...342..598L}.

Indeed, there are a number of common emission phenomena which show
that pulsar magnetospheres can be both stable and dynamic. 
For instance, drifting subpulses are a periodic drift of consecutive single pulses
through the average pulse window, at a rate of tenths to a few times
the spin period \citep{1968Natur.220..231D,2006A&A...445..243W}.
Furthermore, hundreds of pulsars show nulling: a sudden cessation in emission
between otherwise strong and steady pulses, spanning one to a few
spin periods in most cases, but lasting upwards of several hours in
some cases \cite[e.g.][]{2007MNRAS.377.1383W}.  
Extreme nulling events can last for days to years (such sources are sometimes
termed `intermittent pulsars' depending on the arbitrary length of
time for which the pulsar is undetected).  As with shorter nulls, the
transition between the detected emission, `ON', and the null state,
`OFF', occurs in less than 10 seconds, and perhaps within a single
rotation \citep{2006Sci...312..549K}.  For longer-term nulling
pulsars, the spin-down rate during the `ON' state has been detected to be
approximately 1.5- to 2.5-times larger than that during the `OFF'
state \citep{2006Sci...312..549K,2012ApJ...746...63C,2012ApJ...758..141L}.

Mode-changing is a similarly abrupt switch in the pulsed emission
between two (or more) discrete and well-defined pulse profile morphologies
\citep{1971MNRAS.153P..27L,1988ApJ...324.1048R} that 
 are observed to occur over the same time-scales as pulse nulling 
\citep[i.e. seconds to years; e.g.][]{2007MNRAS.377.1383W,2011MNRAS.411.1917W,2014ApJ...780L..31B}.
This is also occasionally identified in variable linear
\citep{1976Natur.263..202B}, and circular polarisation
\citep{1978ApJ...223..961C}. 
In some cases, pulse profile changes are also correlated
with large changes in spin-down rates (0.3--13\%) \citep{2010Sci...329..408L}.

The above emission phenomena are thought to be connected, and related
to changes in the current flows in the pulsar magnetosphere
\citep{2010Sci...329..408L,2012ApJ...752..155V,2012ApJ...746...60L}.
For example, nulls can be regarded as a mode during which emission
ceases or is not practically detectable.  Most pulsars may exhibit these emission
phenomena, but the effects of pulse-to-pulse variability are hard
to identify in sources with low flux densities, because summing over
many pulses is required to achieve sufficient signal-to-noise ratio
  (S/N). Also, it is possible that fewer pulsars are identified with longer
characteristic timescale phenomena, such as mode-changing and extreme nulling, because
frequent observations over a longer timespan are necessary to enable this.
The hope is that better understanding these emission characteristics
may lead to a better understanding of the magnetospheric emission
mechanism itself.

{\psr} is the focus of this work because it has been shown to exhibit a host of
emission phenomena.  This includes bursts of subpulses that drift
slowly in longitude, towards both the leading and, more commonly,
trailing edge of the main pulse \citep{2007A&A...469..607W}.  The
nulling fraction has previously been estimated to be 7(2)\%\footnote{Note: Throughout the paper, numbers in parentheses are the uncertainties
corresponding to the least significant figure in the value quoted.}, and nulls occur clustered
in groups more often than at random
\citep{2009MNRAS.393.1391H,2009MNRAS.395.1529R}. 
Recently, \cite{2012MNRAS.427..114Y} showed that {\psr} nulls over a broader range of timescales
(from minutes to at least five hours), and that the pulsar is detected only
70--90\% of the time when observed.  They also found no detectable
change in spin-down rate between emitting and long null states to a
limit of less than 6\% fractional change.  Long-term timing of the
pulsar shows relatively high timing noise residuals of 89\,ms
\citep{2004MNRAS.353.1311H}, including changes in the sign of the spin
frequency second derivative \citep{2013ApJ...775....2S}.  Thus, {\psr}
is a good candidate for studying how the many facets of the pulsar emission
mechanism are observed and related in one source.  However, admittedly this may complicate matters for
studying any one of these properties individually.

{\psr} was one of the first pulsars to be discovered
\citep{1968IAUC.2100....1C}, and is one of the brightest radio pulsars
in the Northern sky, with 73(13)\,mJy flux density at 400\,MHz
\citep{1995MNRAS.273..411L} and a distance of
0.32$^{\rm{+(8)}}_{\rm{-(5)}}$\,kpc \citep{2012ApJ...755...39V}.    
It has been detected from 40\,MHz \citep{2007ARep...51..615K} up to
32\,GHz \citep{2008A&A...480..623L}, with little profile evolution
\citep{1978A&A....68..361B}.  It is also one of ten slowly-spinning, 
rotation-powered pulsars with pulsations detected in X-rays \citep{2004ApJ...615..908B}.  The interpulse in the
pulse profile suggests that it is nearly an orthogonal rotator, which
is also supported by polarisation observations \citep{2001ApJ...553..341E}.

To further investigate the emission phenomena displayed by {\psr}, we conducted radio
observations at low frequencies ($<$200 MHz) using the Low-Frequency
Array (LOFAR), where brightness modulation due to scintillation is
averaged out across the band. We also conducted simultaneous
observations using the LOFAR, Westerbork, Lovell, and Effelsberg
telescopes to record data from 110\,MHz to 2.7\,GHz. 

The observations and data reduction are described in Section
\ref{sec:obs}.  In Section \ref{sec:res} we report on the data
analysis and results of the observations. We show that the long-term
nulls previously reported are in fact a very weak and sporadically
emitting mode, similar to that identified in PSR~B0826$-$34
\citep{2005MNRAS.356...59E}.  Hereafter we refer to this mode as the
`quiet' (Q) emission mode, and therefore the regular strong emission as `bright' (B),
analogous to other mode-changing pulsars\footnote{{\psr} has a dispersion measure (DM) 
of 19.454(4)\,pc\,cm$^{\rm{-3}}$ \citep{2004MNRAS.353.1311H}, and both diffractive and refractive
scintillation is observed -- the decorrelation bandwidth at 1.7\,GHz is
81(3)\,MHz \citep{2013MNRAS.436.2492D}. We note, however, that the
2--10 hour `disappearances' of {\psr} interpreted as due to
scintillation by \citet{2013MNRAS.436.2492D} are more likely to be instances of the
weakly-emitting Q-mode observed in this work.}. 
In Section \ref{sec:disc} we
discuss these and further results and present our conclusions.

\section{Observations}\label{sec:obs}

LOFAR observations of {\psr} were obtained on 2011 November 13--14, 2012 February 9, and 2013 April 7
 \citep[see][for a description of LOFAR]{2013A&A...556A...2V}.
 Between 5 and 21 High-Band-Antenna (HBA) core stations were coherently combined
using the LOFAR Blue Gene/P correlator/beam-former to form a
tied-array beam \citep[see][for a description of LOFAR's pulsar
observing modes]{2011A&A...530A..80S}.  For a summary of the specifications of the observations, see Table \ref{tb:1}.

Starting at 23:18 UT on 2011 November 13, and ending at 08:12 UT on 2011 November 14, 26 three-minute
observations were taken with a gap of 19\,minutes between
successive observations. Data were taken using the central six Core
stations (the `Superterp'), and were written as 32-bit complex values
for the two orthogonal linear polarisations at a centre frequency of
143\,MHz and bandwidth of 9.6\,MHz. 
{\psr} switched to emitting in the B-mode during the thirteenth observation.
The pulsar was not detected during the other 25 similar observations in this campaign, even after all observations were summed together in time.

To further evaluate the moding behaviour at LOFAR frequencies, a
longer, three-hour observation was taken on 2012 February 9 using five
of the central HBA Core stations. The linear polarisations were
summed in quadrature and the signal intensities (Stokes I) were
written out as 32-bit, 245.76-$\upmu$s samples, at a centre frequency
of 143\,MHz with 47.6\,MHz bandwidth.

To investigate the broadband behaviour of {\psr}, a further eight-hour
observation on 2013 April 7 was conducted simultaneously using LOFAR
and three telescopes observing at higher frequencies: the Westerbork
Synthesis Radio Telescope (WSRT), the Lovell Telescope and the
Effelsberg 100-m Telescope, see Table \ref{tb:1} for a summary.  The
LOFAR observation used 21 HBA Core stations, where the linear polarisations
were summed in quadrature, and the signal intensities (Stokes I) were
written out as 32-bit 245.76-$\upmu$s samples, at a centre frequency of
149\,MHz with 78\,MHz bandwidth.
The WSRT
observation was taken in tied-array mode using the PuMaII pulsar
backend \citep{2008PASP..120..191K}.  Baseband data were recorded for 8
slightly overlapping 10-MHz bands, resulting in a total bandwidth of
71\,MHz centred at 346\,MHz.  The Lovell Telescope observed the pulsar
using the digital Reconfigurable Open Architecture Hardware (ROACH)
backend. Dual polarisations were Nyquist sampled over a
400-MHz band centred at 1532\,MHz and digitised at 8\,bits.  The
Effelsberg 100-m Telescope observation was conducted using the PSRIX backend
(ROACH-board system for online coherent dedispersion) that recorded data at a centre frequency
of 2635\,MHz with 80\,MHz bandwidth. 

The data from the LOFAR observations were converted to 8-bit samples
offline.  Data from all observations were coherently
dedispersed, except the LOFAR observations on 2012 February 9 and 2013 April 7 that were incoherently dedispersed, 
and folded using the pulsar's rotational ephemeris
\citep{2004MNRAS.353.1311H} and the \textsc{dspsr} program
\citep{2011PASA...28....1V}, to produce single-pulse integrations.
Radio frequency interference (RFI) in the pulsar archive files was
removed in affected frequency subbands and time subintegrations
using both automated (\textsc{paz}) and interactive (\textsc{pazi})
programs from the \textsc{PSRCHIVE}\footnote{see
  http://psrchive.sourceforge.net for more information.} library
\citep{2004PASA...21..302H}.  The LOFAR polarisation data were calibrated for
parallactic angle and beam effects by applying a Jones matrix \citep[see][for further details]{2015A&A...576A..62N}. 
The linear polarisation parameters in the LOFAR data from the 2011 November 14 observation
were also corrected for Faraday rotation, see Section \ref{subsec:desc} for further details.

\begin{table*}
\centering
  {

\caption[Summary of LOFAR and simultaneous multi-frequency
  observations of {\psr}.]{ Summary of observations. Columns 1--8
  indicate the observing telescope, date, start time, integration
  time, centre frequency, bandwidth, individual channel width, and
  sampling time.  Column 9 shows the duration of the B-mode
  relative to the start time, where `+' indicates that the pulsar
  remained in the B-mode until at least the end of the observation.  For the
  LOFAR observations on 2011 November 14, 2012 February 9, and 2013 April
  7: the observation IDs are L34789, L45754, L119505, respectively;
  HBA[0,1] stations included in tied-array mode were CS00[2--7],
  CS00[2,3,4,6,7] and
  CS0[01-07,11,17,21,24,26,28,30,32,101,103,201,301,302,401],
  respectively. Due to increased observing
  bandwidth and number of stations used, the 2012 February 9 and 2013 April 7 LOFAR observations were 2 times and 10 times more
  sensitive than the original 2011 November 14 LOFAR observation,
  respectively.     }

\label{tb:1}
\renewcommand\tabcolsep{0.07cm}
  \begin{tabular}{l l l l l l l l l } 
\hline\hline\\[-0.8em]  
Telescope & Date & Start Time & $\tau_{\rm int}$ & Frequency & Bandwidth & Channel-width & $\tau_{\rm samp}$ & Time in B mode\\
 & (dd.mm.yyyy) & (UT) & & (MHz) & (MHz) & (kHz) &   ($\upmu$s) & \\[0.4em] 
\hline\\[-0.8em]  
LOFAR & 14.11.2011 & 03:42 & 3.0 (min)&  143 & 10 & 12 & 81.9 & 1.69+ (min)\\
\hline\\[-0.8em]
LOFAR & 09.02.2012 & 20:30 & 3.0 (h)&  143 & 48 & 12 & 245.8 & 2.71+ (h)\\
\hline\\[-0.8em]  
LOFAR & 07.04.2013 & 14:00 & 8.0 (h)&  149 & 78 &  12 & 245.8 & 3.07--3.11,5.09+ (h)\\
WSRT  & 07.04.2013 & 14:00 & 8.0 (h)&  346 & 71 &  24 & 64.7 &   3.07--3.11,5.09+ (h)\\
Lovell & 07.04.2013 & 14:00 & 8.0 (h)&  1532 & 400 &  4000 & 518.2   & 3.07--3.11,5.09+ (h)\\
Effelsberg& 07.04.2013 & 14:43 & 7.3 (h)& 2635 & 80 & 195 &  259.1  & (3.07--3.11),5.09+ (h)\\
\hline\\[-0.8em]  
  \end{tabular}}
\end{table*}

\section{Analysis and Results}\label{sec:res}

Here we describe the analysis and results of the observations
summarised in Table \ref{tb:1}.  {\psr} switched between emission modes at least once during each observing campaign.  
Initially we focus on the time- and frequency-averaged properties at LOFAR frequencies, Section \ref{subsec:desc}, and at a range of frequencies, Section \ref{subsec:multi}.  In Section \ref{subsec:sp}, we focus on the single-pulse analysis of the data from the simultaneous observations.  

\subsection{Emission Modes Description}\label{subsec:desc}

\begin{figure*}
\begin{center}
\includegraphics[width=0.9\textwidth]{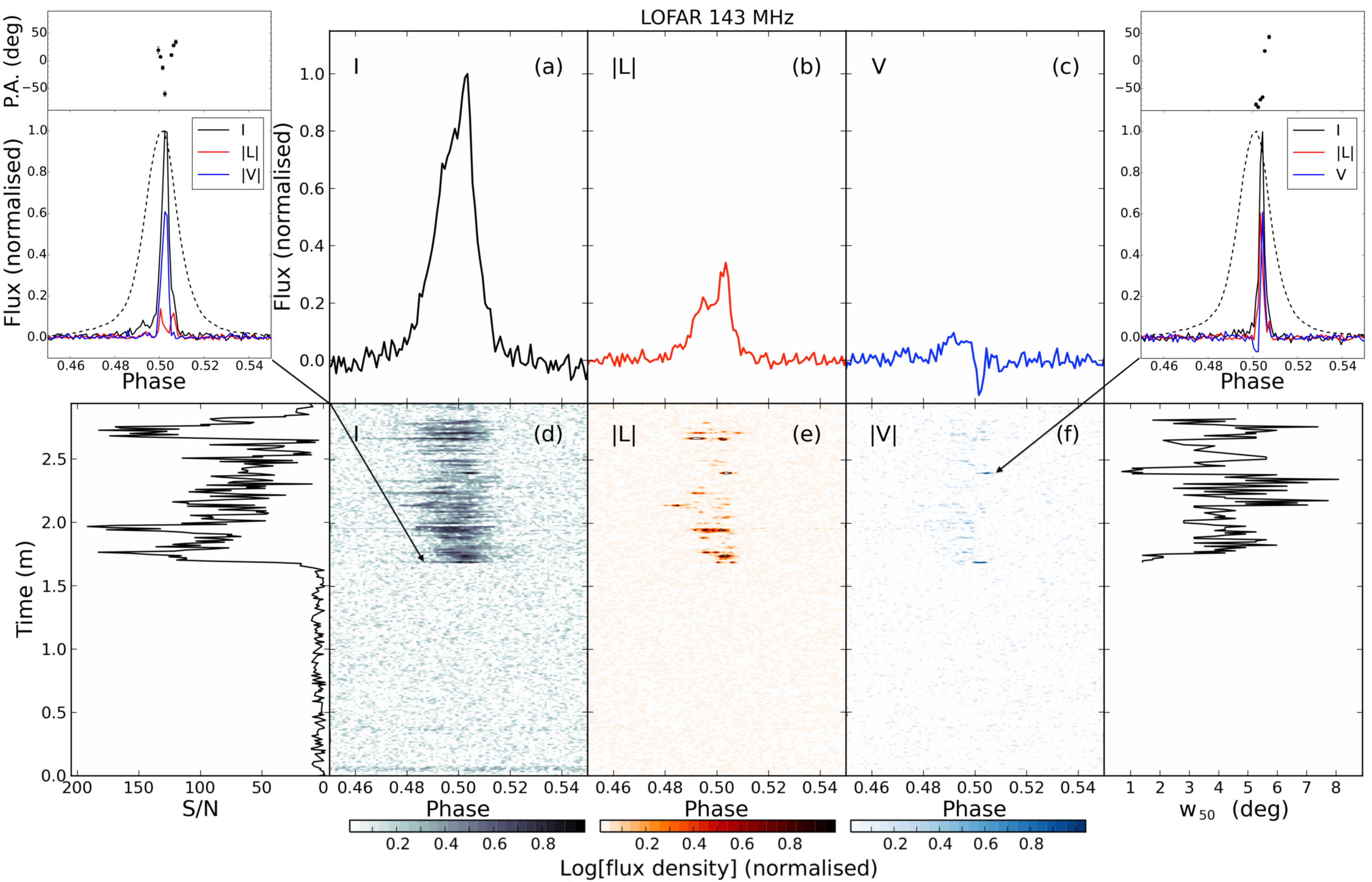}

\caption[Overview of three-minute LOFAR observation on 14 November
  2011]{Overview plot of the LOFAR observation on 2011 November 14.
  \emph{Panels a,b,c:} Time- and frequency-averaged pulse profiles
  for the B-mode in total (Stokes $I$), linearly polarised ($|L|$ = Stokes
  $\sqrt{Q^{\rm2}+U^{\rm2}}$) and circularly polarised (Stokes $V$)
  intensities, respectively. The average profile before the start of the B-mode is not shown as it is
  indistinguishable from the noise in this short observation.
  \emph{Panels d,e,f:} The flux density of single pulses against pulse
  phase and integration time in total (increasing from white to
  black, see colour bars), linearly polarised (increasing from white to red) and absolute
  circularly polarised (increasing from white to blue) intensities,
  respectively. 
   \emph{Lower left:} S/N of total intensity single pulses against time.  \emph{Lower right:}
  Full-width-half-maximum, $w_{\rm{50}}$, against time for total intensity single pulses with S/N$>$7.  \emph{Inset plots: } Normalised
  polarisation profiles for two narrow and highly polarised single
  pulses indicated with arrows: initial B-mode pulse (\emph{left}),
  and second pulse with similar characteristics at 2.4\,min
  (\emph{right}).  \emph{Lower panels} show the total (black),
  linearly polarised (red) and circularly polarised (blue)
  intensities.  The stable average pulse profile in total intensity
  from the 2012 February 9 LOFAR observation is
  also shown (dashed line) for the same frequency and
  using the same timing ephemeris (from Pilia et al. submitted). \emph{Upper panels} show the position angle
  of the linear polarisation (P.A. = 0.5 tan$^{\rm{-1}}(U/Q)$).  }
\label{fig:pre}
\end{center}
\end{figure*}

Figure \ref{fig:pre} shows an overview of the emission characteristics
of {\psr} over the course of the three-minute LOFAR observation taken
on 2011 November 14, during which the pulsar switched to B-mode
emission.  This is the only observation with recorded polarimetric
information.  Figure \ref{fig:pre}, \emph{panel d}, clearly shows the
sudden increase in total flux density 1.69\,minutes after the start of the
observation.  \emph{Panels e} and \emph{f} show that this is also the
case for the polarised flux density. The B-mode continues for at least 160
single pulses until the end of the observation.  This is an inadequate
number of pulses to obtain a stable average pulse profile, which is
shown for comparison in Fig. \ref{fig:pre}, \emph{inset plots}. 
The total duration of the B-mode in this case is not known
exactly. During all 12 preceding and 13 following three-minute
observations no emission was detected, and we infer that {\psr} was
either nulling and/or emitting in the Q-mode.  Assuming this was also the case during the
19-minute gaps between the observations, the duration of the Q-mode and/or nulling
can be constrained to at least 4.4\,h before and 4.75\,h after the
observation summarised in Fig. \ref{fig:pre}.  This multi-hour
duration is similar to the length of prolonged nulls
also observed by \cite{2012MNRAS.427..114Y}. It is also possible that
`flickers' of emission on the order of minutes in duration, such as
that shown in Fig. \ref{fig:pre}, could have occurred in between
observations.

Inspection of the single pulses from Fig. \ref{fig:pre} shows that the transition between the Q and B-modes
is very rapid, occurring within one single rotational period.  This is
notable in the profiles of the single pulses and in the dramatic
increase, by at least 60-times, in median S/N.
We also note significant changes in the pulse width and morphology. Figure \ref{fig:pre}, \emph{lower right},
shows that the full width at half maximum, $w_{\rm{50}}$, of the initial
single pulses in the B-mode are very narrow (1.4$^{\circ}$), compared
to the average pulse profile at the same frequency (7$^{\circ}$, see
Fig. \ref{fig:pre}, \emph{inset plots}, and below). Subsequent B-mode
pulses increase in width and drift towards earlier and later pulse phases until
an equally narrow pulse (1.1$^{\circ}$) occurs 80 pulses later at 2.40\,min.  This may
be expected due to intrinsic variability, e.g., subpulse drifting. 
Moreover, the narrow pulses are located
towards the trailing edge of the average pulse profile, similar to
what is observed from PSR~B1133+16 \citep{2003A&A...407..655K}.

Further investigation of the polarisation properties was also carried out.
The RM of the integrated pulse
profile and the two narrow highly-polarised pulses was determined using
RM-synthesis \citep{2005A&A...441.1217B}. The observed RM of the
integrated B-mode pulse profile was determined to be
6.28(4)\,rad\,m$^{\rm{-2}}$.  We find the RMs of the narrow single
pulses to be consistent with this value.  This RM value was used to correct the 
linear polarisation Stokes parameters for Faraday rotation in order to maximise the linear polarisation in the average pulse profile.
The RM due to the ISM alone was determined to be 5.25(8) rad m$^{\rm{-2}}$,
after correcting for ionospheric Faraday rotation
\citep{2013A&A...552A..58S}.  This is in good agreement with (within
less than 2$\sigma$), and more precise than, the current ATNF pulsar
catalogue value of 5.9(3) rad m$^{\rm{-2}}$
\citep{1974ApJ...188..637M}. The DM and corrected RM values can be used to determine the 
Galactic magnetic field direction and magnitude parallel to the line-of-sight
towards {\psr} \citep[e.g.][]{2008MNRAS.386.1881N}, which was calculated to be 0.332(6)\,$\upmu$G. 
These results are discussed further in Section~\ref{subsec:probe}.

Variable pulse-to-pulse fractional polarisation was also identified. The two narrow 
pulses previously identified show relatively high polarisation (especially circular) fractions, see
Fig. \ref{fig:pre}, \emph{inset plots}. The initial B-mode pulse is
55(5)\% right-hand circularly polarised, and the narrow pulse at 2.4\,min
is 56(5)\% left-hand circularly polarised.  The initial B-mode pulse and subsequent
narrow pulse are 10(5)\% and 38(5)\% linearly polarised, respectively.  The
fractional circular polarisation for these single-pulses is therefore
considerably larger than for the average pulse profile, which is 6(5)\%
circularly polarised and 15(5)\% linearly polarised.

\begin{figure}
\begin{center}
\includegraphics[width=0.49\textwidth]{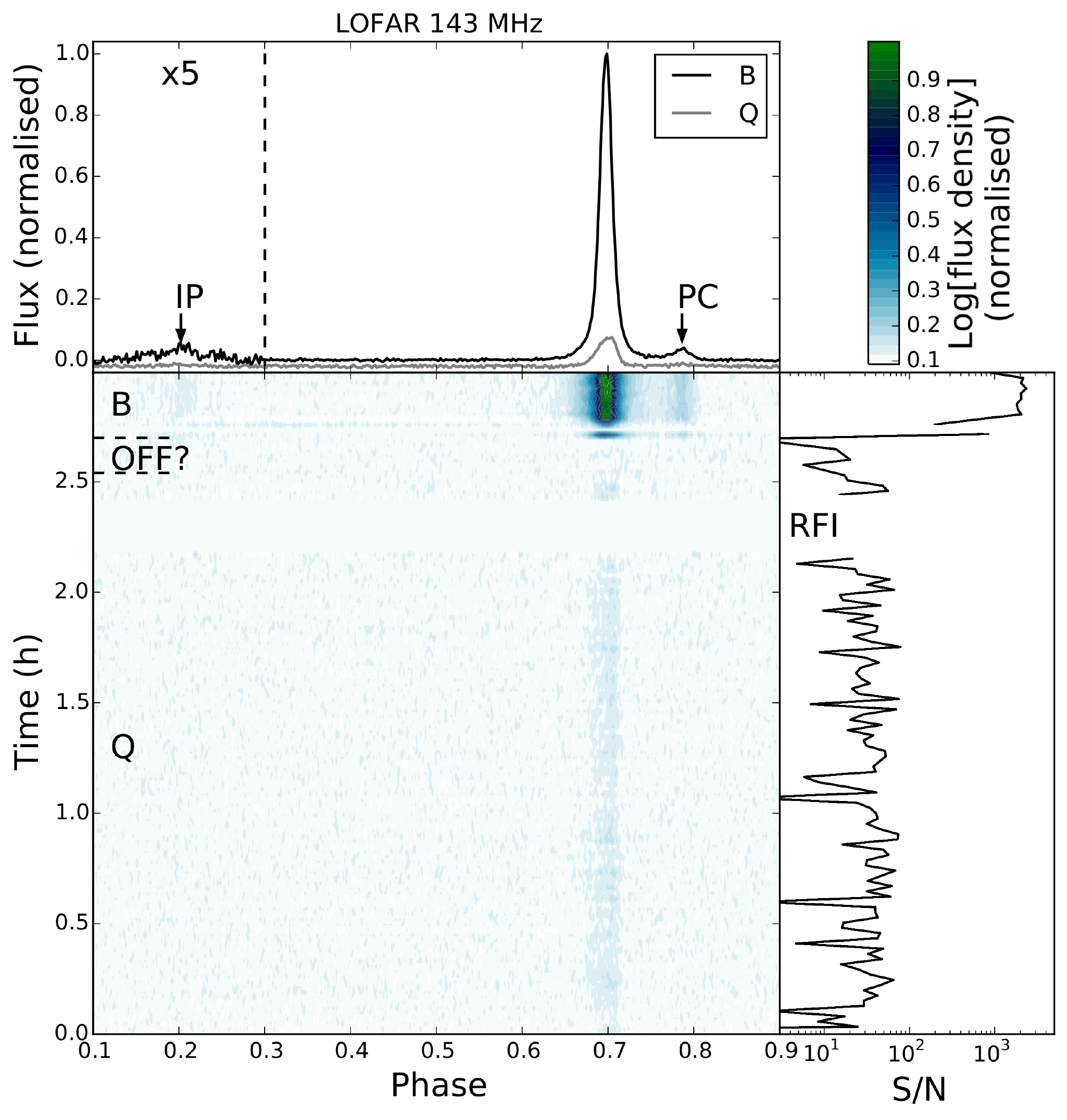}

\caption[Overview of three-hour LOFAR observation on 9 February
  2012]{Overview plot of the LOFAR observation taken on 2012 February 9.
    \emph{Lower left panel:} Flux density of the pulsar against pulse phase and integration time 
  (logarithmic scaling, increasing from white through blue to green, see colour bar at the \emph{upper right}). 
  The modes of emission are identified: Q-mode (0--2.55\,h),  B-mode (2.71+\,h), and, in the period of approximately 10 minutes in which
  the pulsar is not detected, tentatively `OFF?' (2.55--2.71\,h). These emission modes are delimited by the dashed lines.
  Subintegrations removed due to RFI are left blank, and the most affected time period is also labelled. 
   \emph{Upper panel:} Time- and frequency-averaged total
  intensity pulse profiles separated by emission mode into Q (0--2.55\,h, grey line) and B (2.71+\,h, black line), normalised to the peak flux of the B-mode
  profile.  The postcursor (PC) and interpulse (IP, 5$\times$  magnified) are also indicated.  
   \emph{Lower right panel:} S/N of 42-second total intensity sub-integrations against integration time.}

\label{fig:lofarobs}
\end{center}
\end{figure}

In order to further examine the moding behaviour, we observed {\psr}
again for a continuous three-hour period using LOFAR on 2012 February 9. As shown
in Fig. \ref{fig:lofarobs}, \emph{lower panels}, 
the pulsar switched from Q- to B-mode emission
after 2.71\,h.  The \emph{lower panels} of Fig. \ref{fig:lofarobs}
clearly demonstrate that weak emission is detected from the start to
2.5\,h into the observation, at approximately the same location in
pulse phase as that of the main pulse during the B-mode.  This weak, but
significant, Q-mode emission has not previously been detected;
this mode was previously thought to be a long-term null. Between the
weak Q-mode at 2.55\,h and the beginning of the B-mode at 2.71\,h, the
flux density of the pulsar decreases even further and no emission is detected.  {\psr} rapidly
transitions to B-mode at 2.71\,h, again within one single rotational period, 
and continues in this mode until the end of the observation.
This is clear from the abrupt increase in S/N, see Fig. \ref{fig:lofarobs}, \emph{lower right}, and also in the pulse profile,
see Fig. \ref{fig:lofarobs}, \emph{upper panel}.  The ratio between the
median S/N of single pulses in the B and Q-modes is approximately 170.

Figure \ref{fig:lofarobs}, \emph{upper panel}, illustrates the average
pulse profiles of the Q and B-modes.  The B-mode pulse profile shows
the known components from the literature, i.e., a narrow main pulse
(MP), postcursor (PC), and interpulse (IP, separated by almost exactly
half a rotation period from the MP)
\citep[see][]{1973NPhS..243...77B}.  The newly discovered Q-mode pulse
profile is notably weaker.  The location of the peak in flux density
is also skewed towards slightly later pulse phases compared to the
B-mode MP (+2(1)$^{\circ}$, 2.9 ms).  Table \ref{tb:2} includes a summary of the
relative intensity, width and locations of the B and Q-mode pulse
profile components. Details of how these were calculated are included
in Section \ref{subsec:multi}.

\subsection{Multi-frequency Analysis}\label{subsec:multi}

\begin{figure*}
\begin{center}
\includegraphics[width=0.445\textwidth]{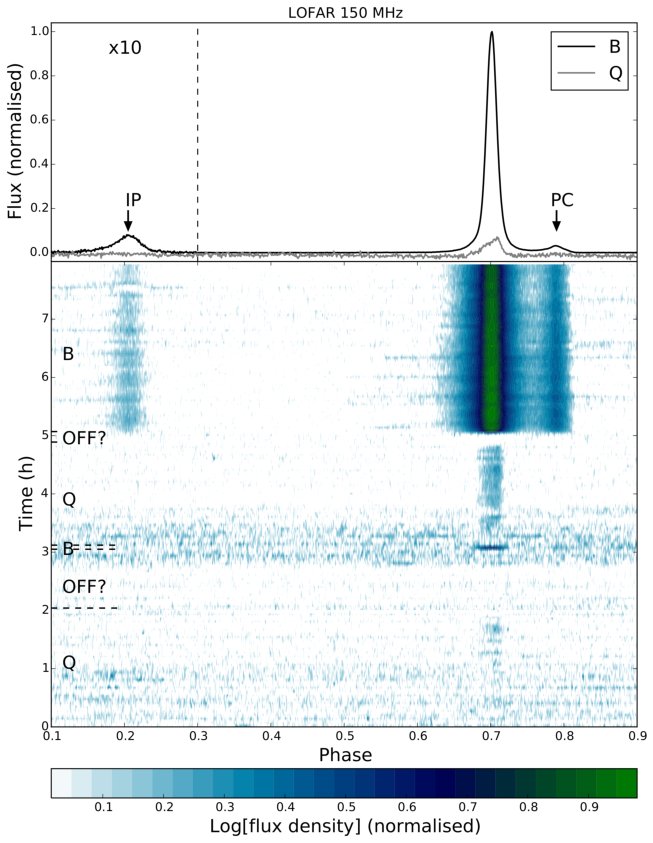}
\includegraphics[width=0.445\textwidth]{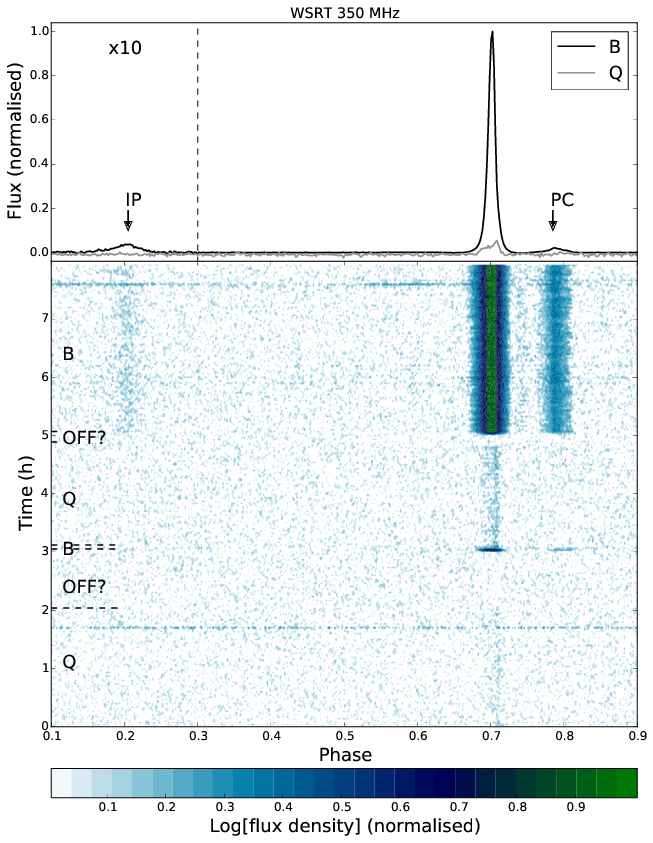}
\includegraphics[width=0.445\textwidth]{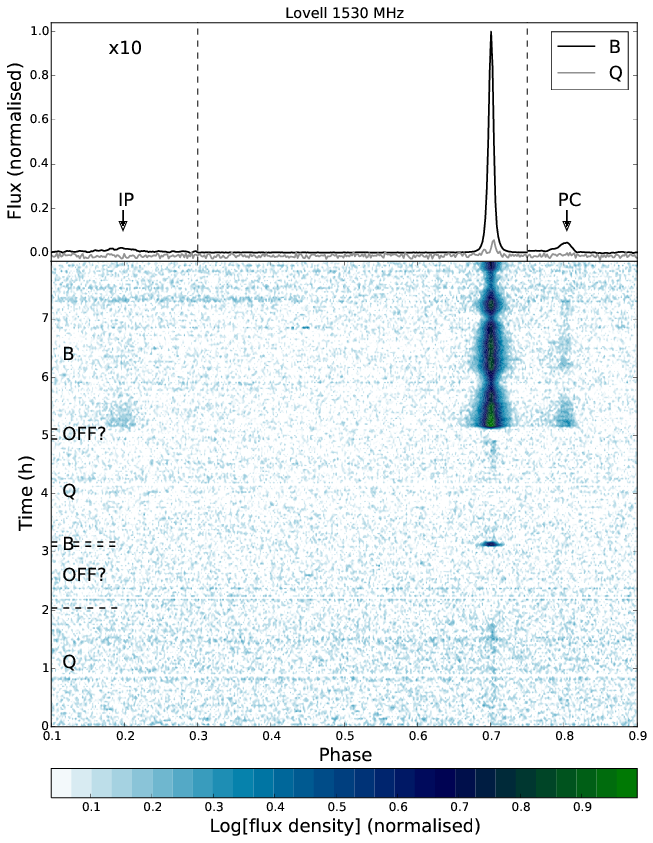}
\includegraphics[width=0.445\textwidth]{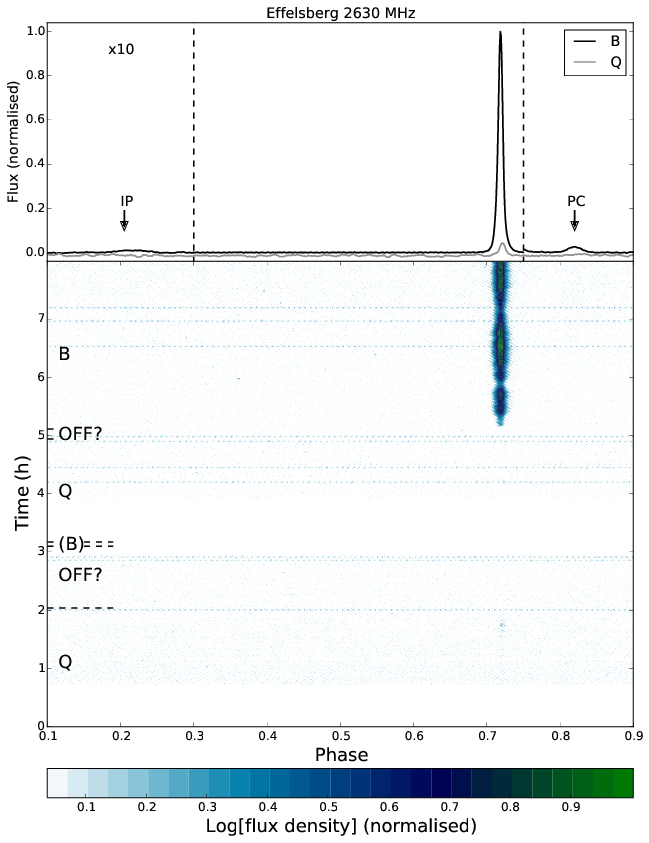}

\caption[Overview of simultaneous multi-frequency observations of
  {\psr} on 2013 April 7] {Overview of the multi-frequency
  simultaneous observations taken on 2013 April 7, using: LOFAR
  (\emph{upper left}), WSRT (\emph{upper right}), Lovell (\emph{lower
    left}), and Effelsberg 100-m (\emph{lower right}).  \emph{Upper
    panels:} Time- and frequency-averaged total intensity pulse
  profiles separated into Q-mode (0--3.07\,h, 3.11--5.09\,h; 10$\times$
  magnified; grey lines) and B-mode (3.07--3.11\,h, 5.09+\,h, black lines), normalised to the
  peak flux of the B-mode profile. There also appear to be periods in which
  the pulsar is not detected (`OFF?'; $\sim$2--3.07\,h, 4.8--5.09\,h). The PC (10$\times$ magnified in the 
  lower plots) and IP (10$\times$ magnified) are also indicated.
  \emph{Lower panels:} Flux density of the pulsar against pulse phase
  and integration time (increasing from white through blue to
  green, see colour bars). Subintegrations removed due to RFI are shown as white. 
  The modes of emission identified (Q-, B-, and, tentatively, `OFF?') are also labelled, and these are delimited by the dashed lines.
   The off-pulse noise level in the LOFAR observation shows an increase around the transition from Q- to B- to Q-mode emission around 3\,h because there was a period of
increased RFI around this time, and more conservative RFI flagging was
done to conserve as much data around this period as possible. 
  Modulation of the B-mode brightness due to scintillation is also visible in the higher frequency Lovell and Effelsberg data.
 }

\label{fig:multiobs}
\end{center}
\end{figure*}

In light of previous studies of other pulsars
\citep[e.g.][]{2013Sci...339..436H}, mode-switching is
expected to occur over a broad-range of wavelengths, even up to
X-rays.  To investigate whether this is the case for {\psr}, we
observed the pulsar using four telescopes (LOFAR, WSRT, Lovell,
Effelsberg) at a range of centre frequencies (149, 345, 1532,
2635\,MHz) simultaneously. The overview of these eight-hour
observations, Fig. \ref{fig:multiobs}, shows that both Q and B-modes
were detected again, and that the behaviour of {\psr} is very similar
across the range of frequencies.  

\begin{figure}
\begin{center}
\includegraphics[width=0.48\textwidth]{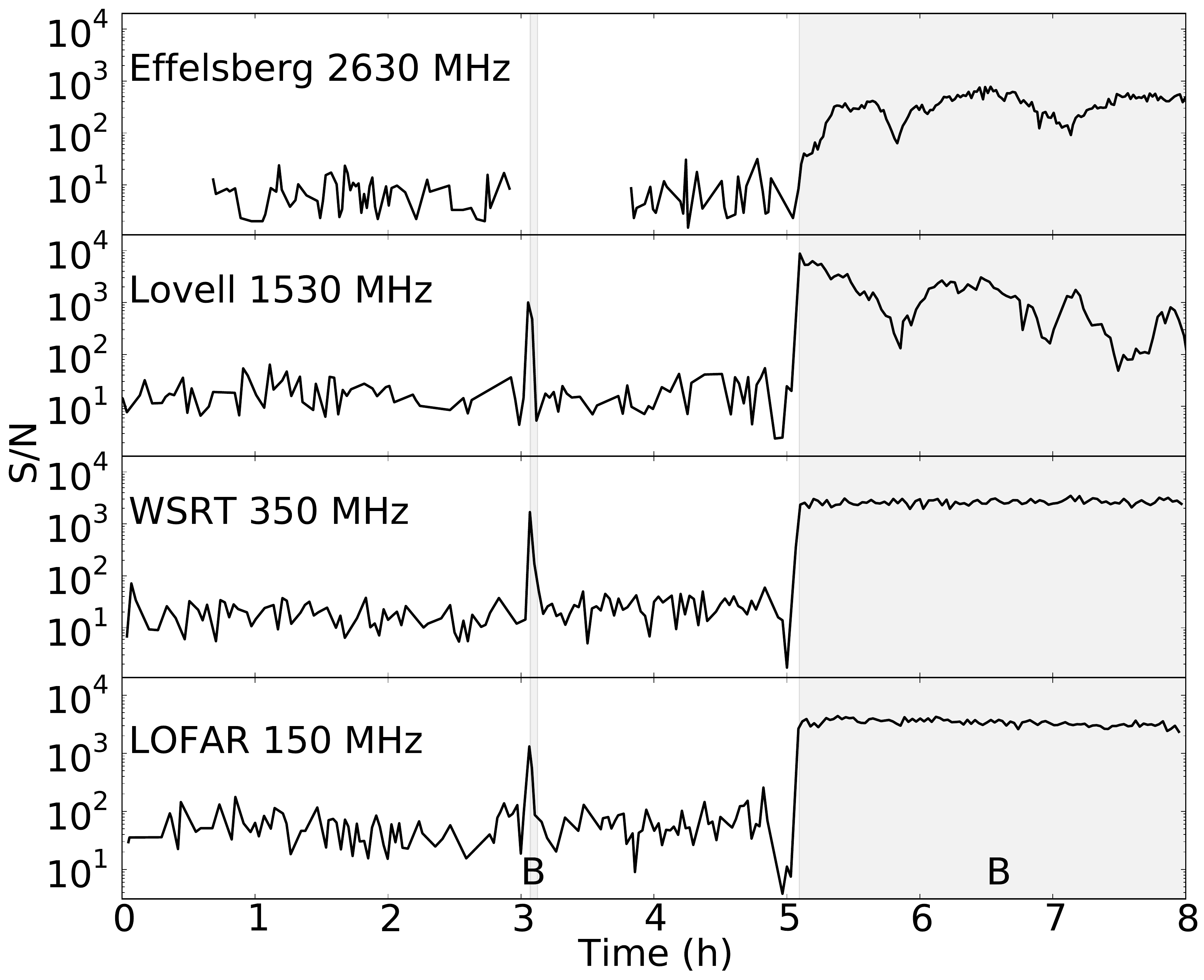}

\caption[S/N of {\psr} during the eight-hour multi-frequency
  observations on 2013 April 7] {S/N of total
  intensity pulses versus integration time from the multi-frequency
  simultaneous observations of {\psr}. \emph{Upper to lower panels:}
  Effelsberg 100-m at 2635\,MHz using 80-s subintegrations; Lovell at
  1532\,MHz using 140-s subintegrations; WSRT at 345\,MHz using 140-s
  subintegrations; LOFAR at 149\,MHz using 110-s integrations.
  Discontinuation of lines indicates flagged subintegrations due to
  RFI.  The periods of B-mode emission are indicated by the shaded grey backgrounds and labelled `B'.}

\label{fig:snr}
\end{center}
\end{figure}

The lower panels in
Fig. \ref{fig:multiobs} show that from the beginning of the
observations until approximately 2\,h, the weak Q-mode emission is
visible at all four frequencies.  Between $\sim$2 and 3.07\,h the Q-mode
becomes even more weak and is practically undetectable. After
this, the pulsar abruptly `flickers' into B mode emission for
approximately 160 pulses.  The pulsar transitions back into the
weakly-emitting Q-mode between 3.11 and 4.8\,h, which is most clearly
detected at the two lower frequencies, but also towards the end of
this period at the two higher frequencies. The flux in the Q-mode once
again decreases even further and becomes practically undetectable
between 4.8 and 5.09\,h, just before the pulsar transitions again into
B-mode for the remainder of the observation.  This is also
demonstrated in terms of S/N against observing time in
Fig. \ref{fig:snr}.

Figure \ref{fig:multiobs}, \emph{upper panels}, illustrate the B-mode
profiles and the previously unknown and much weaker Q-mode pulse
profiles between 110\,MHz and 2.7\,GHz.  The B-mode MP, PC, and IP are
visible at all frequencies and, as expected from the literature, there
is little profile evolution across the observed frequency range.  Again,
the Q-mode is much weaker compared to the MP of the B mode, and the
average peak flux occurs towards slightly later pulse phases
($>+$1$^{\circ}$, depending on frequency). 

On all occasions where the emission switches from the Q to B-mode, the
transition occurs within one single rotational period at all
frequencies, identified by the notable increase (70$\times$) in S/N of
single pulses and change in pulse profile morphology.  Figure \ref{fig:multiobs},
\emph{lower panels}, also show that the mode switch from Q to B at 5.1\,h
occurs simultaneously for all pulse profile components, i.e., the MP,
PC and IP, at all frequencies.

To determine whether the mode-switch occurs coincidently across all
frequencies observed, the MJD of the initial B-mode single-pulse after
the mode change was determined for each frequency. The expected time
delays of equivalent pulses due to dispersion in the cold plasma of the ISM between
the centre frequencies used at Effelsberg, Lovell, and WSRT and the centre frequency
used at LOFAR are 3.63, 3.61, and 2.97\,s, respectively.  Taking this into account, we
find that the mode-switch is coincident at the frequencies observed to
within 0.01\,s.

On the single occasion where we observe the pulsar switch from the B
to Q-mode, it is more difficult to pinpoint the transition
because the flux density of single pulses decreases gradually over a few
rotational periods, similar to the behaviour during the 2011 November 14 observation shown in Fig. \ref{fig:pre},
\emph{panel d}, rather than showing an abrupt change in S/N. For the three lower frequencies, B-mode emission was detected over 
a 160 single-pulse timescale during the B-mode `flicker' at 3.1\,h.  Since the transition
from the Q to B-mode was determined to be simultaneous across the
observed band, it is probable that this is also the case for the
B-to-Q-mode change.

Figure \ref{fig:snr} shows the S/N against time for the simultaneous
observations. This clearly demonstrates the simultaneous wide-band
transitions between emission modes, especially the decrease in flux
density in the Q-mode just before the rapid transition to the B-mode
at 5.1\,h.  Here we again see that the B- and Q-mode differ in median flux density by over two orders-of-magnitude.

Fig. \ref{fig:snr} illustrates fluctuations in pulse total intensity due to scintillation, which are
much more evident in the prolonged B-mode emission at the two higher
observing frequencies. This is because the scintillation bandwidth
becomes comparable to the observing bandwidth at the higher
frequencies. 
The DM of {\psr}, measured using the LOFAR observation on 2013 April 7 using the \textsc{PSRCHIVE pdmp}
program \citep{2004PASA...21..302H}, was found to be 19.475(3)\,pc\,cm$^{\rm{-3}}$.
This amount of dispersion contributes to {\psr} falling into the strong
diffractive scintillation regime when observed below approximately 5\,GHz \citep{2013MNRAS.436.2492D}.  
At lower frequencies the scintillation bandwidth
is much smaller, causing its effects to be averaged out more
effectively across the band.

The B- and Q-mode multi-frequency average pulse profiles obtained from
the 2012 November 9 and 2013 April 7 observations, see
Figs. \ref{fig:lofarobs} and \ref{fig:multiobs}, were used to compare
the behaviour of the pulse profile components with frequency.  The profile
components were fitted with Gaussian functions using the
Levenberg-Marquardt algorithm \citep{LevenbergLeastSquares} to provide
the normalised peak amplitude, peak location, $w_{\rm 50}$, and full width at tenth of maximum, $w_{\rm 10}$, values, and their respective errors.
Table \ref{tb:2} provides the summary of these characteristics for the pulse profile components at
multiple frequencies for B and Q-mode emission.  Peak amplitude is measured in 
percentage relative to the B-mode MP.

After quantifying characteristics of the profile features at each
frequency listed in Table \ref{tb:2}, the trend with respect to
frequency was explored.  Assuming that each feature, $x$, follows a
power law with respect to frequency, $x\propto\nu^{\rm \eta}$, a
weighted best fit was performed to determine the appropriate value for
the exponent, $\eta$. The
resulting values and errors for the exponents are shown in the bottom
row of Table \ref{tb:2}.

\begin{table*}
\centering
  {
\caption[Summary of results from the observations of {\psr}, including
  pulse component widths and relative intensities]{Summary of the
  pulse profile components from the LOFAR observation on 2012 November
  9 (LOFAR$_{\rm 1}$) and the simultaneous multi-frequency
  observations on 2013 April 7.  Columns 1 and 2 indicate the
  telescope and central frequency used.  $w_{\rm {50}}$ of the B-mode
  MP, PC, and IP, and the Q-mode pulse are shown in columns 3, 7, 11,
  and 14, respectively.  $w_{\rm {10}}$ of the B-mode MP and PC are
  summarised in columns 4, and 8, respectively.  The peak flux
  location of the B-mode PC and IP, and the Q-mode pulse, with
  respect to that of the B-mode MP (where `+' indicates later pulse phases) 
  are shown in columns 5, 9, and 12,
  respectively.  The relative amplitude of the peak flux of the B-mode
  PC and IP, and Q-mode pulse, with respect to the MP (corrected for
  different integration times using Table \ref{tb:1}) are summarised
  in columns 6, 10, and 13, respectively. The bottom row shows the
  frequency dependence of each of the profile features as the exponent
  of a power law, $\eta$.}

\label{tb:2}
\hspace*{-0.5cm}
\renewcommand\tabcolsep{0.07cm}
  \begin{tabular}{| ll | ll |  llll  | lll  |  lll |} 
\hline\hline\\[-0.8em]  
 &  & \multicolumn{2}{|l|}{B-mode MP} & \multicolumn{4}{|c|}{B-mode PC} & \multicolumn{3}{|c|}{B-mode IP} & \multicolumn{3}{|c|}{Q-mode} \\
\cmidrule(r){3-4} \cmidrule(r){5-8} \cmidrule(r){9-11} \cmidrule(r){12-14}
Telescope & Frequency      & $w_{\rm 50}$ &  $w_{\rm 10}$ &location& amplitude & $w_{\rm 50}$ & $w_{\rm 10}$ &  location & amplitude & $w_{\rm 50}$  & location & amplitude & $w_{\rm 50}$ \\
(Name)  & (MHz)         & ($^{\circ}$) &  ($^{\circ}$)   &  ($^{\circ}$)  &   (\%)         & ($^{\circ}$) &      ($^{\circ}$) &    ($^{\circ}$)  &   (\%)        & ($^{\circ}$)  &     ($^{\circ}$)     &    (\%)      & ($^{\circ}$)  \\ 
\hline\\[-0.8em]  
LOFAR$_{\rm 1}$ & 143 & 6.68(5) & 14.06(8) & +31.5(1) & 3.76(5) & 9.8(3) & 17.9(5) & --178.5(7) & 0.90(5) &  20.5(6) & +2(1) & 0.89(5) & 9.4(2) \\
LOFAR                 & 149 & 6.73(3) & 13.36(8) & +31.5(1) & 3.19(5) & 9.6(2) & 17.4(2) & --179.3(2) & 0.75(5) & 16.2(5)  & +1.9(2) & 0.62(5) & 9.6(4)\\
WSRT                  & 346 & 4.29(3) & 9.14(8) & +31.7(1) & 2.09(5) & 9.4(2) & 17.2(3) & --180.0(3) & 0.30(5) & 15.2(2) & +1.8(2) & 0.47(3) & 8.7(5)\\
Lovell                   & 1532  & 3.12(3) & 6.33(8)   & +36.0(2) & 0.69(4) &  8.3(2) & 15.2(2) & --181(1)  & 0.23(6) & 12.3(8) & +1.3(5) & 0.39(9) & 3.7(3) \\
Effelsberg           & 2635  & 2.89(7) & 5.97(9) & +36.1(1) & 0.51(3) & 7.2(3) & 13.2(3) & --182(1)  & 0.12(3) & 11(1) & +0.9(6) & 0.7(3) & 1.8(4) \\
\hline\\[-0.8em] 
All                       & $\eta$ & --0.33(4) & --0.30(4) & +0.050(9) & --0.66(6) & --0.08(2) & --0.08(2) & +0.005(1) & --0.63(2) & --0.19(6) & --0.16(4) & --0.3(2) & --0.37(9) \\
\hline\\[-0.8em]  
  \end{tabular}}
  \end{table*}
  
We find that the width of all pulse profile components decrease with
increasing frequency in both emission modes, although this trend is
most notable for the Q-mode pulse.  This is consistent with previous
results for the B-mode pulse profile
\citep[e.g.][]{1994A&AS..107..527K}, and also with radius-to-frequency
mapping \citep[RFM; e.g.][]{2002ApJ...565..500G}.  
The largest intra-channel DM smearing at the lowest LOFAR frequency accounts for 
0.3$^{\circ}$ in pulse longitude, which does not affect the conclusions obtained from the low-frequency pulse profile.
The B-mode PC and IP move further from the MP and decrease in
amplitude relative to the MP with increasing frequency. The positions
of the PC and IP and their frequency dependence are in agreement with
previous results \citep{1986ApJ...304..256H}.  This is further discussed
below in Section \ref{sec:disc}.  Table \ref{tb:2} shows that the
amplitude of the PC has the strongest frequency dependence and the
location of the IP is the least dependent on frequency.

The relative pulse amplitude of the Q-mode profile is approximately equal to the amplitude of the B-mode IP, 
and also decreases with increasing frequency.  
The peak amplitude of the Q-mode is located towards later pulse phases compared
to the B-mode MP at the same frequencies, and seems to arrive progressively earlier in pulse
phase towards higher frequencies.  Furthermore, Table \ref{tb:2} shows that 
both B- and Q-mode pulse profiles show no
significant change over one year between the LOFAR observations from 2012
February 9 and 2013 April 7.

\begin{figure}
\begin{center}
\includegraphics[width=0.48\textwidth]{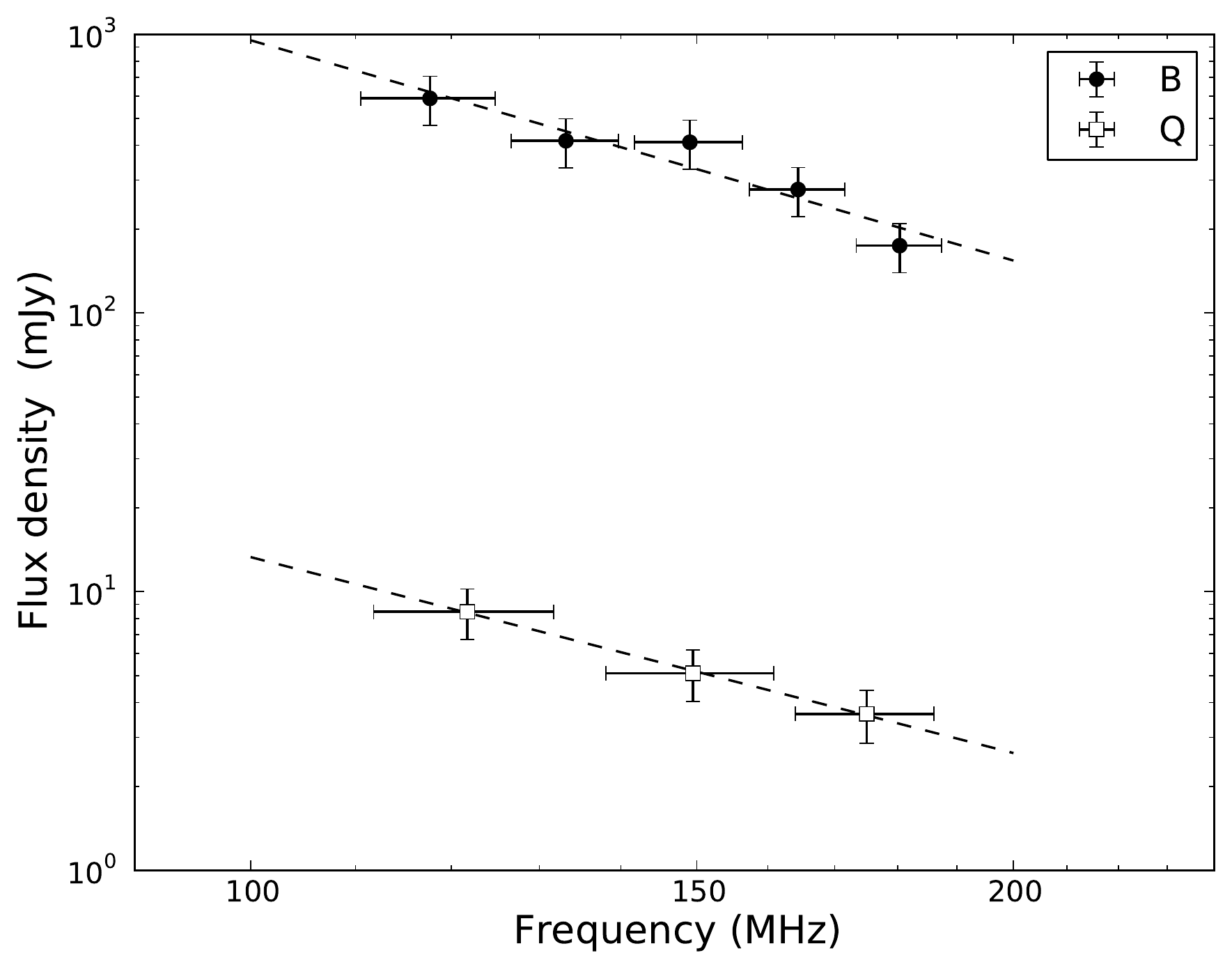}

\caption{Low-frequency spectra of {\psr}, separated into B- (filled
  circles) and Q-mode emission (open squares), obtained from the LOFAR
  observation taken on 2013 April 7.  Frequency-axis error bars show
  the bandwidth used to calculate each point.  The dashed lines show
  the best power-law fit to the data.}

\label{fig:spectra}
\end{center}
\end{figure}

To further investigate the relative properties of the B and Q-modes at
low-frequency, the spectrum in each mode was calculated using the
eight-hour LOFAR observation.  The flux density, $S$, as a function of
frequency was obtained using the radiometer equation, including an
empirically-derived and tested formula for the effective area of the
LOFAR stations, and correction for the primary beam using the
theoretically-derived and tested beam model, described in Hassall,
et al. (in prep.), and references therein.  The B-mode and Q-mode data were divided into 5 and 3 
subbands, respectively, deemed to be appropriate to optimise the respective S/N.  
A line of best fit was calculated assuming a single power
law across the bandwidth of the observation, $S \propto \nu^{\alpha}$, and the error was determined
considering the flux density errors only. 
Figure \ref{fig:spectra} shows the resulting spectra in the two
emission modes.

The difference in flux density between the B and Q-modes is striking, see
Fig. \ref{fig:spectra}. In fact, the average flux density in the
B-mode, 370\,mJy at 149\,MHz, is almost two orders-of-magnitude
greater than that of the Q-mode, 5\,mJy.  This is probably a
factor in the previous non-detection of this weak Q-mode.  

The exponent, $\alpha$, obtained from the continuous power-law best fits, weighted by the uncertainties,
for the B and Q-modes are --2.6(5) and --2.3(1), respectively.
Therefore, there is no significant difference in the power-law index
of the spectra between emission modes.   This is further discussed in
Section \ref{sec:disc}. The relatively steep spectral index and strong scintillation are also 
possible factors in the previous non-detection of the weak Q-mode emission at higher frequencies.
The increased bandwidth used for the observation using the Lovell Telescope on 2013 April 7 in this work (400\,MHz) 
also provided almost a factor of two better sensitivity in comparison to the observations in \citet{2012MNRAS.427..114Y}.  

\subsection{Single-pulse analysis}\label{subsec:sp}

Potentially interesting emission features were investigated and compared for the B
and Q-modes by analysing single pulses from the eight-hour simultaneous
observations. This was conducted using a software package provided by P. Weltevrede (private communication)
to construct centred, gated, de-baselined, and further RFI-excised single-pulse
archives \citep[e.g.][]{2006A&A...458..269W}.  This enabled us to extract information regarding nulling,
subpulse drifting, and pulse energy distributions (PEDs).

\subsubsection{Pulse Energy Distributions}

\begin{figure*}
  \begin{center}
 
  \includegraphics[width=1\textwidth]{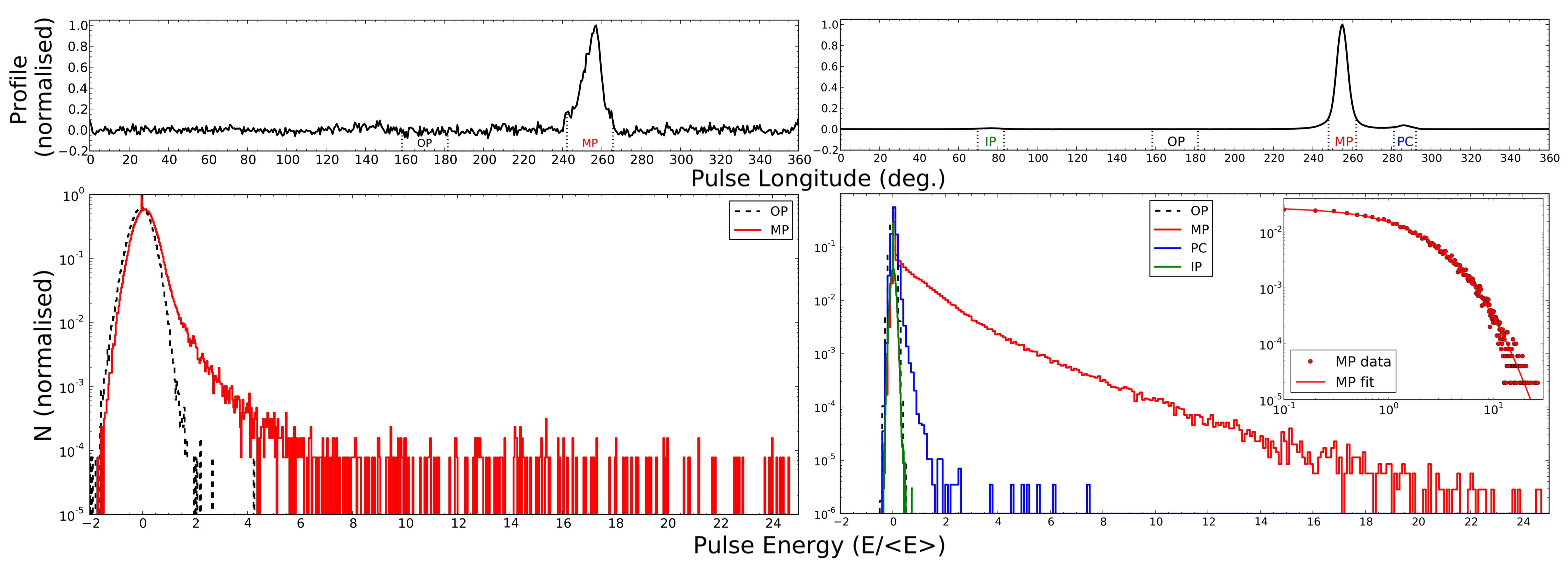}
 
  \caption[Longitude-resolved pulse energy distributions separated by
    emission mode]{ \emph{Upper panels:} Normalised average pulse
    profiles (black line) for the Q-mode (\emph{left}) and B-mode
    (\emph{right}), showing longitude ranges included in the pulse-energy analysis
    (delimited by the black dotted lines).  \emph{Lower
      panels:} Normalised longitude-resolved PEDs for the Q- 
    (\emph{left}) and B-mode emission (\emph{right}), using the LOFAR
    observation taken on 2013 April 7.  The MP, PC, IP, and off-pulse (OP)
    distributions are shown (red, blue, green, and black dotted lines,
    respectively).  \emph{Inset panel:} B-mode MP PED data (red points) and 
    best lognormal fit to the data (red line), plotted on logarithmic axes.  }

\label{fig:LRPEhist}
  \end{center}
  \end{figure*}

Longitude-resolved PEDs were calculated for each pulse profile
component of the B and Q-modes by integrating over a selected pulse longitude bin range, see Fig. \ref{fig:LRPEhist}.  
Pulse longitude ranges corresponding to $w_{\rm{10}}$ were selected to optimise S/N.
These are indicated underneath the average pulse profiles in Fig. \ref{fig:LRPEhist}, \emph{upper panels}.
The longitude range of the off-pulse (OP) region
shown in Fig.  \ref{fig:LRPEhist} was selected to correspond to the
range selected for the Q-mode pulse, and to be located mid-way between
the B-mode MP and IP.

Figure \ref{fig:LRPEhist}, \emph{upper panels}, shows the
normalised flux-density profiles of the B and Q-mode.  Note the difference in 
the off-pulse noise level.  The lower panels
in Fig. \ref{fig:LRPEhist} illustrate the normalised PED with respect
to the average pulse energy, $<$E$>$, shown separately for the Q
(\emph{left}) and B-modes (\emph{right}). The B-mode MP and PC PEDs
are clearly distinct from that of the off-pulse region and resemble lognormal
probability density distributions, see \emph{inset panel}. Using a method similar to that in \citet{2006A&A...458..269W}, the best lognormal fit
to the PED for the MP for pulse energies 1.5--20 times greater than the average
yields a standard deviation of 0.77(15). The IP PED is comparable to
that found for the off-pulse region because of its relatively weak
single-pulse flux.

Q-mode emission shows a distinctly different
PED, which falls off much more rapidly at low energies and resembles a
power-law distribution.  The Q-mode PED is less distinguishable from
that of the off-pulse region at lower pulse energies due to the much
weaker emission, but shows a tail of less frequent higher-energy
pulses. The best power-law fit to the PED for pulse energies
1--8-times greater than the average gives an exponent of --3.2(2).

To investigate how the energy of the pulses affects the average pulse profile,   
we constructed integrated PEDs from the longitude-resolved PEDs, shown in
Fig. \ref{fig:LRPEhist}, to obtain the average pulse profile for a number of pulse energy ranges.
We find that the most energetic B-mode single
pulse profiles show PCs located towards lagging pulse phases.  Moreover, these pulses are
approximately eight-times brighter compared to the average pulse
profile.  Despite this, the average pulse profile shows remarkably
little pulse profile evolution with increasing pulse energy, similar
to that observed in PSR~B1133+16 \citep{2003A&A...407..655K}. In addition, single pulses with the
highest flux in the Q-mode have smaller $w_{\rm 50}$ values than the
average pulse profile. There is also a more notable contrast in pulse
profile evolution with energy in the Q-mode.  That is, pulses with energies
greater than twice the average possess profiles akin to a skewed
Gaussian, similar to that in Fig. \ref{fig:LRPEhist}, \emph{top
  right}.  Pulses with energies less than twice the average have
profiles that are weak (discernible from the baseline at approximately
3-$\sigma$), irregular, and somewhat top-hat-like in shape, for examples see Fig. \ref{fig:spQ}, \emph{left}. 
Hence, the pulse profiles in the Q-mode seem to show much greater variability with energy 
compared to those in the B-mode.

\subsubsection{Nulling}

The integrated PEDs were used for complimentary analysis of the
nulling fraction (NF). We estimated the NFs by subtracting the off-pulse
pulse-energy samples from the on-pulse PEDs, as described in 
\citet{2007MNRAS.377.1383W}. The analysis was undertaken separately
for the Q-mode, the 160-pulse B-mode `flicker', and the prolonged
B-mode from the 7 April 2013 LOFAR observation, summarised in
Fig. \ref{fig:multiobs}.  

{\psr} displays nulling in both emission
modes.  More specifically, the NF for the prolonged B-mode was determined to be
1.8(5)\%. This is lower than, but still within 2.2-$\sigma$ of, previous estimates
\citep{2009MNRAS.393.1391H}.  The NF for
the B-mode `flicker' was found to be 15(1)\% -- almost ten times
larger than that during the prolonged B-mode emission.  

Further inspection of the single pulses during this short-lived period of
B-mode emission also indicates that the NF increases towards the
transition back to the Q-mode. Inspection of simultaneous pulses from
the 346\,MHz and 1.5\,GHz data also indicate that the nulls occur
simultaneously at all frequencies, which is consistent with mode-switching, 
and is observed in several other cases \citep[e.g.][]{1992ApJ...394..574B,2013Sci...339..436H}.
The NF throughout both occurrences of the Q-mode during the
observation was determined to be 80(9)\%. This indicates that there are over
forty-times more nulls during the Q-mode than during B-mode emission.
 However, we note that the apparent nulls (in both B- and Q-mode emission) may be yet another weaker emission state that is currently undetectable.

\begin{figure}
\begin{center}
\includegraphics[width=0.23\textwidth]{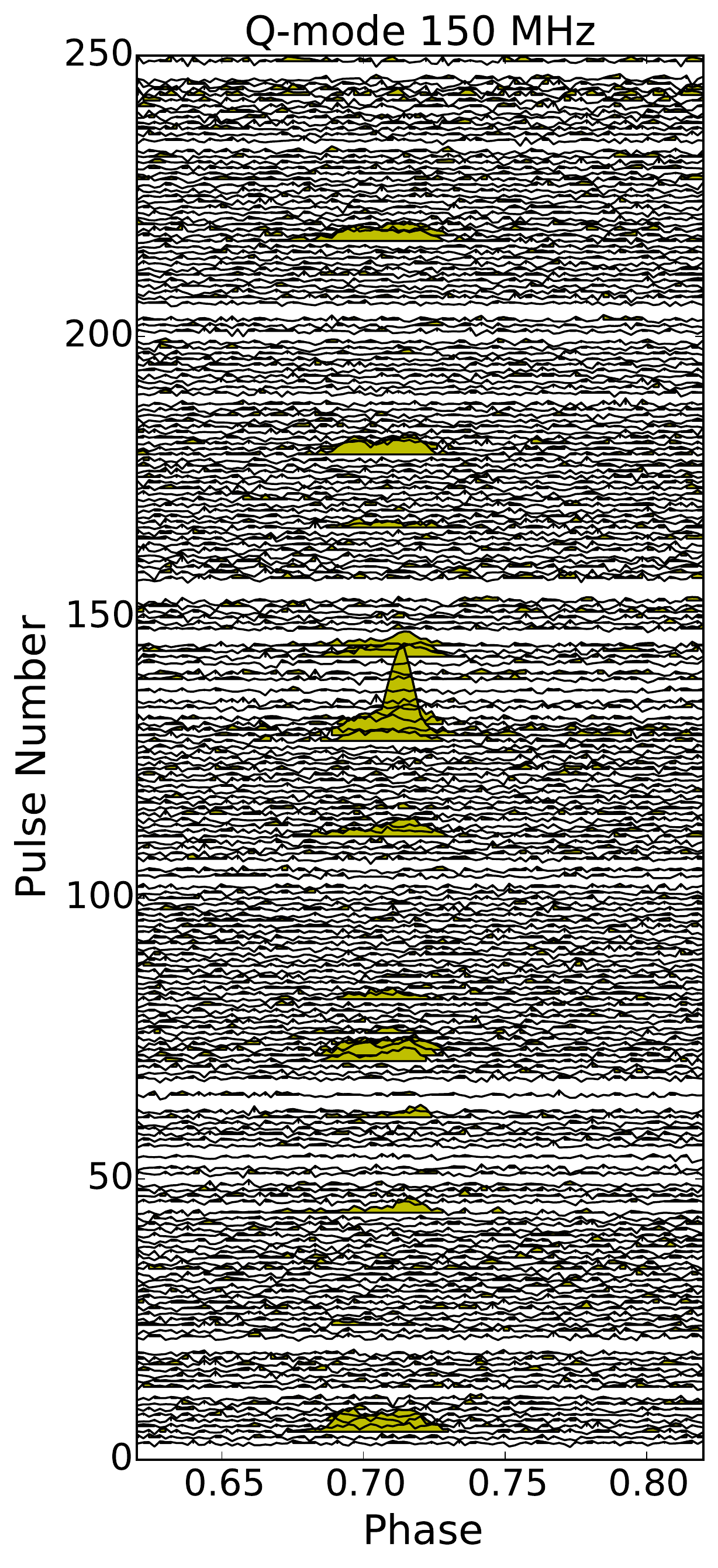}
\includegraphics[width=0.23\textwidth]{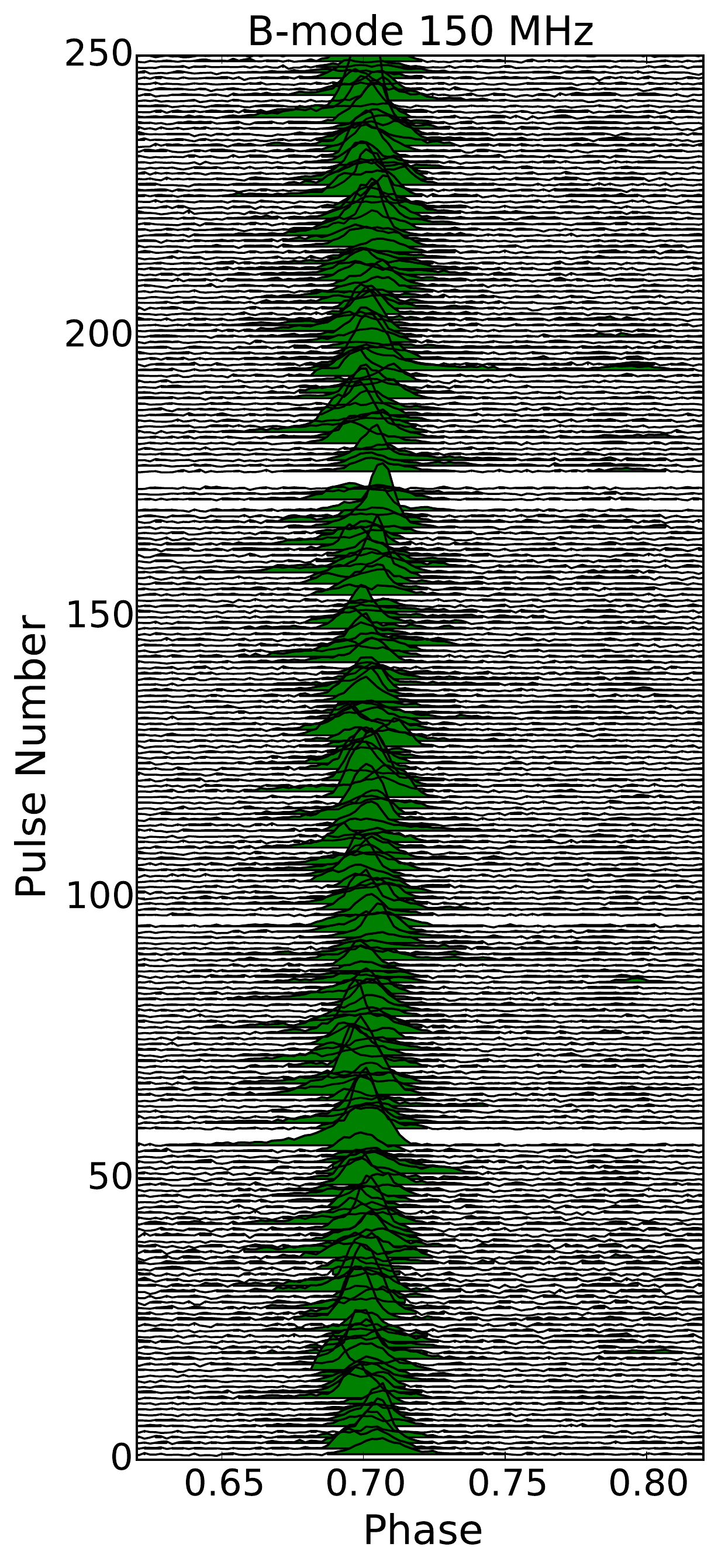}

\caption[Single pulse emission observed from periods of Q-mode
  emission at 149\,MHz]{Representative examples of single-pulse profiles from the
  LOFAR observation on 2013 April 7, arbitrarily normalised. \emph{Left:} sporadic and weak emission during the Q-mode, 
  \emph{right:} B-mode regular emission for comparison.
  Single pulses rejected due to RFI are not  shown. }

\label{fig:spQ}
\end{center}
\end{figure}

Figure \ref{fig:spQ}, \emph{left}, demonstrates the nature of the Q-mode
emission through a 250-single-pulse extract from 4.6\,h into the
observation. A pulse extract from the B-mode emission is also shown in Fig. \ref{fig:spQ}, \emph{right}, for comparative purposes.  
The weak Q-mode emission consists of sporadic, mainly
low-intensity bursts which occur in groups of approximately one to
five pulses with no immediately identifiable periodicity.  Therefore
{\psr} seems to display similar behaviour to the weak-mode emission
identified in PSR~B0826--34 \citep{2005MNRAS.356...59E}, also somewhat
resembling emission from rotating radio transients (RRATs)
\citep{2006Natur.439..817M}.  This is also further discussed in
Section \ref{sec:disc}.

\subsubsection{Subpulse Modulation Features}

\begin{figure*}
\begin{center}
\includegraphics[width=0.34\textwidth]{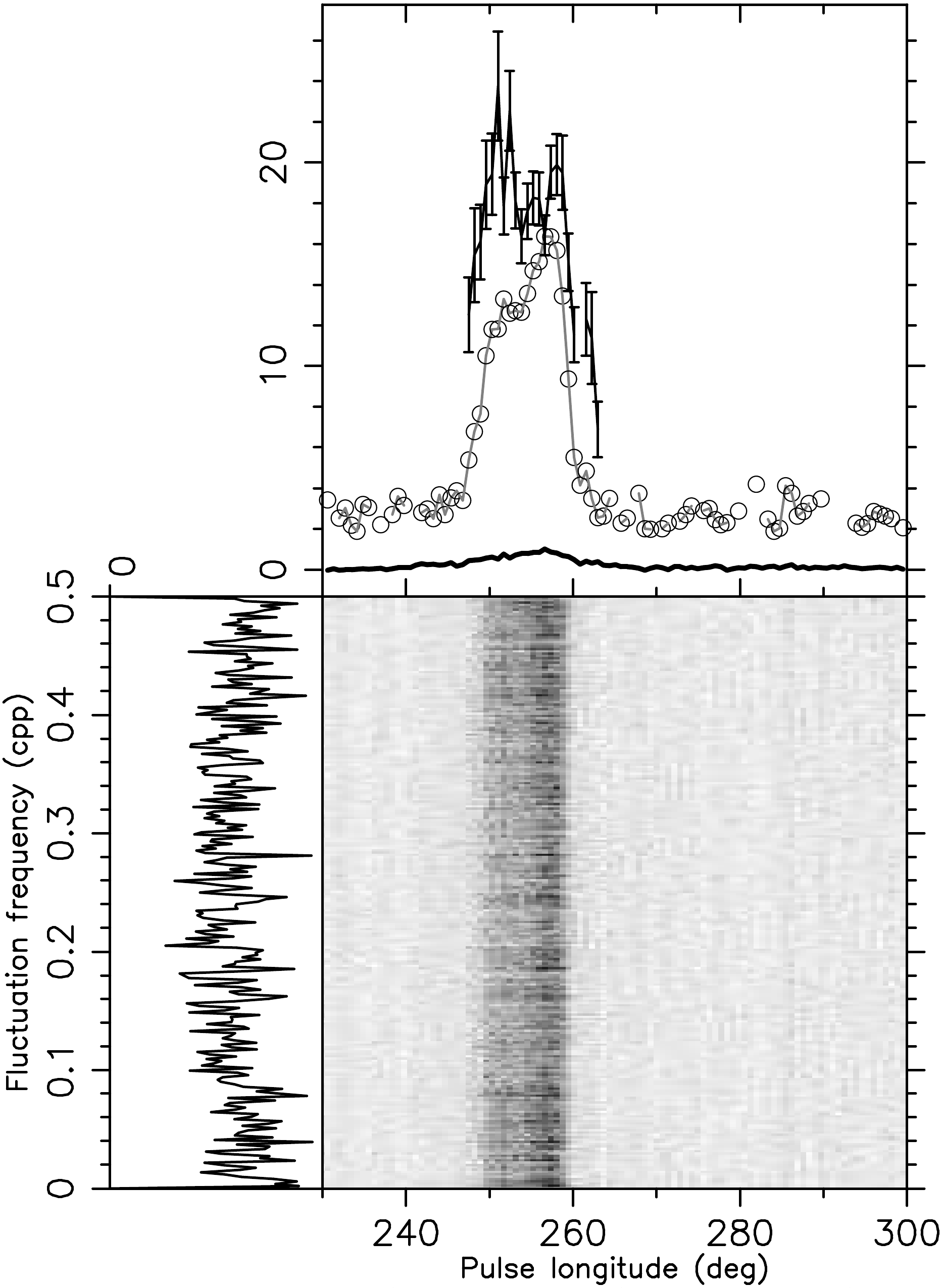}\hspace{15mm}
\includegraphics[width=0.34\textwidth]{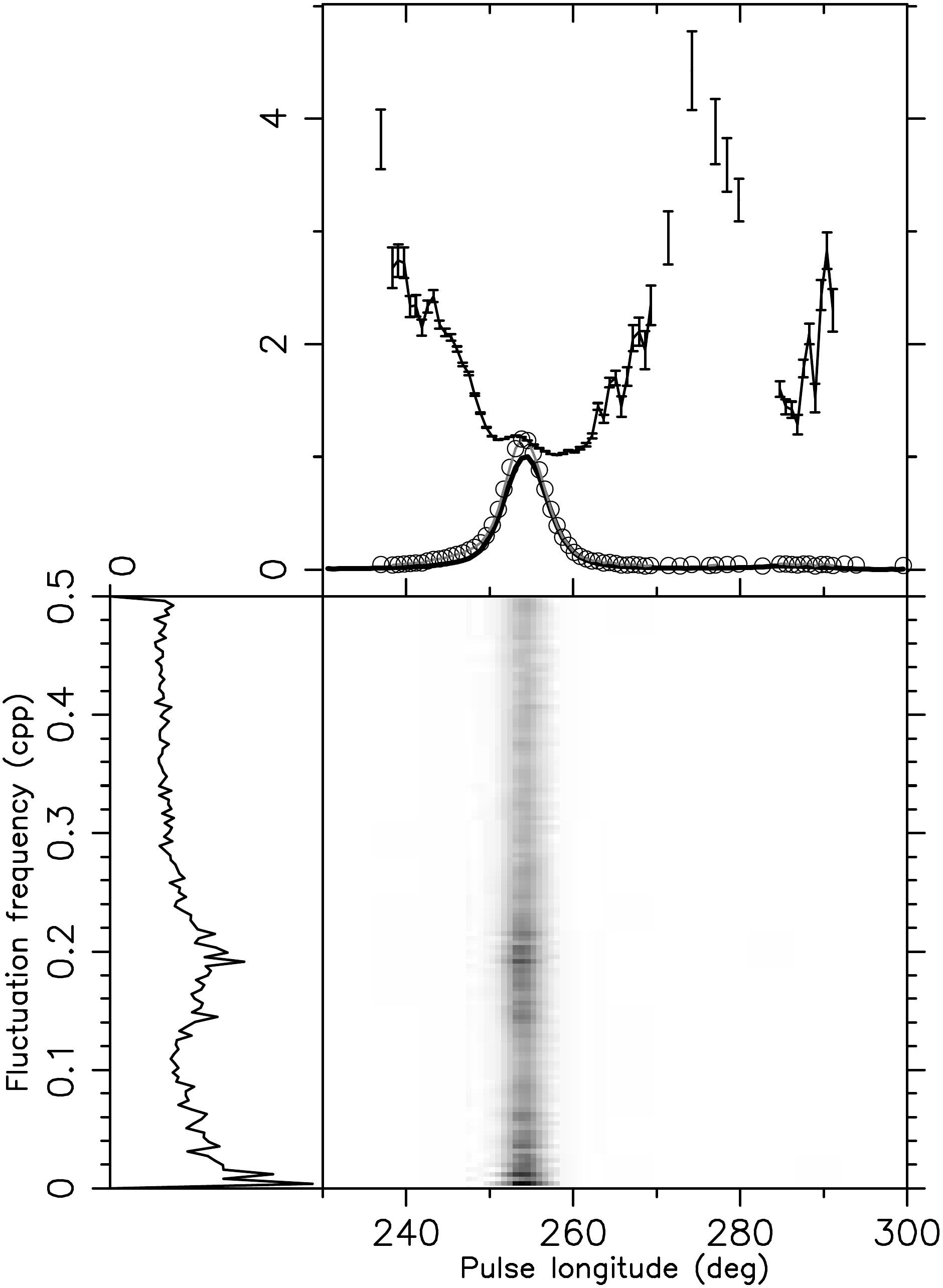}

\caption[Single pulse emission observed from periods of Q-mode
  emission at 149\,MHz]{Pulse variability analysis products from the
  LOFAR observation on 2013 April 7, for the Q-mode (\emph{left}) and the B-mode  (\emph{right}). 
  \emph{Upper panels:} the normalised integrated pulse profile (solid line), the longitude-resolved standard deviation (open circles), and the longitude-resolved modulation index (dimensionless quantity shown as the solid line with error bars). \emph{Lower panels:} the LRFS is shown, with its horizontal integration to the left. }
\label{fig:lrfs}
\end{center}
\end{figure*}

The subpulse modulation properties of the prolonged B-mode at low
frequency were investigated using the LOFAR observation on 2013 April 7, and the method described in
\citet{2006A&A...445..243W}. The aim was to compare the modulation patterns exhibited by the different modes \citep[e.g.][]{2012ApJ...752..155V}.
Pulse stacks, for example see Fig. \ref{fig:spQ}, were used to obtain
the Longitude Resolved Fluctuation Spectrum
\citep[LRFS;][]{1970Natur.227..692B} and hence the vertical separation
between drift bands, $P_{\rm 3}$, measured in pulse periods, $P_{\rm
  0}$.  The LRFS obtained from the B-mode emission, Fig. \ref{fig:lrfs}, \emph{right}, shows two drift
features that are broader than 0.05 cycles per period (CPP).  One
feature is located at the alias border $P_{\rm0}$/$P_{\rm3}$ = 0, and the other feature relates to $P_{\rm
  3}$ = 5.26(5)$P_{\rm 0}$.  Since $P_{\rm 3}$ is generally
independent of observing frequency, this is in excellent agreement
with the value of 5.3(1)$P_{\rm0}$ previously determined at 92-cm
\citep{2007A&A...469..607W}.

We also constructed Two-Dimensional Fluctuation Spectra
\citep[2DFS;][]{2002A&A...393..733E} to determine the horizontal
separation between the drift bands, $P_{\rm 2}$, measured in pulse
longitude. For the B-mode emission we find that although the feature
in the 2DFS peaks at zero CPP, the wings are asymmetric, indicating
that drifting subpulses travel towards later pulse phases
more often.  Calculating the centroid of the asymmetric wings at 20\% of the maximum 
allowed us to extract the value of $P_{\rm  2}$ = +90(20)$^{\circ}$. This is in good agreement with the value
obtained at 92-cm, +70$^{\circ}$$^{+10}_{-12}$
\citet{2007A&A...469..607W}.  Unlike $P_{\rm 3}$, $P_{\rm 2}$ is found
to vary with observing frequency \citep{2003A&A...410..961E}.  Since
$P_{\rm 2}$ = +55$^{\circ}$$^{+40}_{-7}$ at 21-cm
\citep{2006A&A...445..243W}, the value determined here also seems to
be in agreement with the general trend that $P_{\rm 2}$ decreases with
increasing observing frequency for {\psr}.

The same subpulse modulation analysis was also conducted for the
prolonged Q-mode emission from the 2013 April 7 LOFAR observation.  No
significant feature was detected in either the LRFS, Fig. \ref{fig:lrfs}, \emph{left}, or 2DFS, 
indicating that there is variability on all fluctuation frequencies. This may be attributed to
the greater nulling fraction and very weak single pulses. However, PSR~B0943+10, which shares many emission
characteristics with {\psr}, also shows subpulse drifting during
the B-mode and not in Q-mode, which is also comparatively `disordered'
\citep{2011MNRAS.418.1736B}.

The longitude-resolved modulation index, a measure of the factor by
which the intensity varies from pulse to pulse, was derived from the
LRFS \citep{2006A&A...445..243W}, see Fig \ref{fig:lrfs}, \emph{upper panels}.  The median modulation index of the
B-mode MP was found to be 1.4(5). The median modulation index of the PC component was found to be
slightly larger, 1.6(9).  The B-mode MP and PC components show similar longitude-resolved modulation characteristics,
although the PC shows slightly more variability than the MP.
The median modulation index of the Q-mode pulse was found to be over ten-times
larger than the B-mode MP, 17(5), and shows somewhat different longitude-resolved modulation
compared to the B-mode, see Fig \ref{fig:lrfs}, \emph{upper panels}.

The pulse profile stability during B-mode was investigated using the 7
April 2013 simultaneous LOFAR and WSRT data.  The Pearson
product-moment correlation coefficient, $\rho$, was calculated between
an analytic profile of {\psr} and the pulse profile integrated from an
increasing number of effective single pulse numbers, $N_{\rm{efc}}$, as
described in \citet{2012MNRAS.420..361L}.  We find that
towards higher effective pulse numbers ($>$200--500) the LOFAR and
WSRT data follow the expected trend, i.e., 1$-\rho \propto$
$N_{\rm{efc}}^{\rm{-1}}$ \citep{2011MNRAS.417.2916L}.  However, there
are some notable deviations from this trend, which tend to
coincide with narrow, bright pulses.  This suggests that over many
pulse periods the pulse profile is stable, but over short periods the
emission shows intrinsic pulse-to-pulse shape variations.  This is not
unexpected due to previous results that show the array of emission
phenomena exhibited by {\psr}.

\section{Discussion and conclusions}\label{sec:disc}

In this work we have confirmed that {\psr} shows a host of emission characteristics over a wide range of
timescales.  The most surprising finding from the work presented here
is that the long-term nulls, recently found by
\cite{2012MNRAS.427..114Y}, are in fact a very weak and sporadically
emitting mode, which we refer to as Q-mode.  We find that this mode is
over 100 times weaker than the B-mode.  We also
find evidence for a further decrease in flux just before the switch to
B-mode, see Figs. \ref{fig:lofarobs} and \ref{fig:multiobs}.  It is possible that this could be a third emission mode, whereby 
the flux density of emission decreases even further or emission completely ceases.
It would also be quite interesting if this weakest (possibly completely off) mode is
found to always occur before the start of B-mode -- as it
has in the three instances we have observed.  

Considering the
many studies which have been conducted on {\psr}, it is surprising that the long-term
nulling was only recently discovered \citep{2012MNRAS.427..114Y}. It
may be that the emission from {\psr} has changed over a multi-year
timescale, such that mode changing is more frequent or longer in
duration than in previous years, though that remains to be shown.  If
so, this may be due to a multidecadal change in emission, perhaps related to that
observed in the Crab pulsar due to magnetic field evolution \citep{2013Sci...342..598L}.

\subsection{Magnetospheric Switches}

Upon investigation of the newly-discovered Q-mode pulse profile at frequencies between
110\,MHz and 2.7\,GHz, we find the pulse profile comprises a single component located within the
regular B-mode MP envelope, but with a peak located towards
slightly later pulse longitude, and a lower flux density by
approximately two orders of magnitude.  The PC and IP that are present
during the B-mode are not detected during the Q-mode.  
A converse example of a pulsar that exhibits mode-changing, PSR~B0943+10, has a pre-cursor in the Q-mode 
and not in the B-mode at 320\,MHz  \citep{2013Sci...339..436H}.

During the Q-mode, the emission is weak and sporadic,
with a very high NF and modulation index.  This behaviour seems
similar to the Q-mode identified in PSR~B0834--26
\citep{2005MNRAS.356...59E}.  It also seems reminiscent of RRAT-like
emission that may imply a common, or at least related, emission
mechanism. This has previously been proposed from both theoretical
\citep[e.g.][]{2013MNRAS.431.2756J} and observational perspectives 
\citep[e.g.][]{2010MNRAS.402..855B,2011MNRAS.415.3065K}.  The detection of the
weakly-emitting Q-mode from {\psr} could mean that other pulsars
currently identified as extreme-nullers may also show similar weak
emission if higher-sensitivity observations can be done \citep{2005MNRAS.356...59E,2007MNRAS.377.1383W,2014MNRAS.442.2519Y}.

During the simultaneous observation on 2013 April 7,
the initial `OFF?' period lasted approximately 0.87\,h,
and preceded the short-duration 160 single-pulse B-mode `flicker'.
Later in the same observation, and also in the observation on
2012 February 9, the `OFF?' period was much shorter in
duration, and the subsequent B-mode emission was much longer in
duration. This further decrease in radio emission, which tentatively
may be considered as a further separate emission mode, could be caused
by a change in the particle density within the magnetosphere that soon
after causes the switch to B-mode. We also speculate that the duration of the undetected emission towards the
end of the Q-mode, just before the mode-switch to the B-mode, may be
inversely correlated with the length of time the pulsar emits in the subsequent
B-mode. We stress that this is speculative
due to the limited sample, and more work will be done in future to investigate this
by using longer and more frequent low-frequency observations.

Single-pulse analysis allowed us to determine that the transition from the Q-mode to the B-mode
occurs within one rotational period, and that it is broadband
over the radio frequencies observed, similar to other mode changing
pulsars \citep[e.g. PSR B0943+10,][]{2013Sci...339..436H,2014A&A...572A..52B}, and nulling pulsars \citep[e.g. PSR
  B0809+74,][]{1981A&A....93...85B}.  Although, considering the tentative `OFF?' mode just before the 
switch to B-mode emission mentioned previously, the underlying mode change mechanism may be more complex than is 
evident from the single-pulse data.
Within the theoretical framework of
force-free magnetospheres, it has been shown that multiple stable
solutions with different structures are possible, e.g., with different
sizes of closed field line region
\citep{2005A&A...442..579C,2010MNRAS.408L..41T}.  Therefore, the B and
Q-modes could be caused by two stable magnetospheric states with
different magnetospheric structures, affecting the current of
relativistic particles, and therefore the broadband radio emission
mechanism.  However, the mechanism that generates the wide range of
observed timescales (e.g. each emission state lasting many hours
and mode-switches occurring within one rotational period) is still elusive.

\subsection{Considering Spin-down}

{\psr} shows no detectable change in spin-down rate between modes
\citep{2012MNRAS.427..114Y}, similar to other pulsars which exhibit
mode-changing on similar timescales, e.g., PSRs B0834--26 and B0943+10, 
whilst other extreme-nulling pulsars identified as
`intermittent' often do
\citep{2006Sci...312..549K,2012ApJ...746...63C,2012ApJ...758..141L}. In
part, this is because it is difficult to detect small changes in
spin-down if transitions occur more frequently than for the more
extreme nulling pulsars \citep{2012MNRAS.427..114Y}.  This may also be
due to the electromagnetic spin-down torque exerted on the neutron
star, proportional to
$\mathrm{sin^{\rm2}\theta+(1-\kappa)cos^{\rm2}\theta}$
\citep{2006ApJ...643.1139C}, where $\theta$ is the alignment angle
between the spin axis and magnetic moment, and $\kappa$ the ratio between the angular velocity of the pulsar and the 
angular velocity at the death line.  Namely, if there is some small change in
$\theta$ between emission modes, then the change in spin-down torque
is less severe in cases of nearly orthogonal rotators
\citep[e.g. {\psr},][]{2001ApJ...553..341E}.  This is less certain in cases of nearly aligned rotators \citep[e.g. PSR
  B0943+10,][]{2001MNRAS.322..438D} \citep{2012ApJ...746L..24L}.

\subsection{Emission Characteristics}

Although we find that the spectral indices of the B and Q-modes are
consistent, within errors, the broad-band spectrum of {\psr} shown in
Hassall et al. (in prep.) indicates a possible spectral turn-over at
127(25)\,MHz.  Future observations using the LOFAR Low Band Antennas
(30--90\,MHz) would be ideal for further investigating the spectral
properties of {\psr} in both emission modes (if the Q-mode is
detectable), including providing constraints on the presence of a
spectral turn-over, and hence supplying more clues about the origin of
the difference between emission modes.

{\psr} also shows certain emission characteristics which are
reminiscent of several other pulsars.  The PCs associated with the
highest energy single pulses that are 8-times brighter than average
are reminiscent of the Vela pulsar's (PSR~B0833$-$45) `bump' component
after the main pulse, where `giant micro-pulses' with flux densities
exceeding 10-times that of the mean were observed
\citep{2002MNRAS.334..523K}.
For comparison to the highly polarised single pulses detected in the 2011 November 14 LOFAR observation, 
PSR B0656+14, which 
may be a nearby RRAT, 
also shows strong highly-polarised single pulses 
limited to the leading and central regions of the average pulse profile \citep{2006A&A...458..269W}. 
Moreover, narrow pulses observed after the
transition to the B-mode occur towards the trailing edge of the
average pulse profile, similar to that observed in PSR~B1133+16
\citep{2003A&A...407..655K}.  The presence of an IP located close to 180 degrees pulse longitude from the MP, a highly
linearly polarised PC \citep[][]{1999ApJS..121..171W,2015A&A...576A..62N} in the pulse profile, and B- and Q-mode emission
is also reminiscent of PSR~B1822$-$09 \citep{2010MNRAS.404...30B,2012MNRAS.427..180L}.

Polarisation observations have allowed 
the MP to be classified as a core component and the IP to be classified as a
possible core component of the opposite magnetic pole \citep[][and references therein]{1999ApJS..121..171W}.
The PC's high linear polarisation fraction could originate from induced
scattering of the main pulse emission into the background by the
particles of the ultra-relativistic highly magnetised plasma
\citep{2008MNRAS.384L...1P}. This could also explain the trend in the
PC location growing more distant with increasing frequency, which is
contradictory to the behaviour expected from RFM \citep[e.g.][]{2002ApJ...577..322M} and has
lead to reluctance in its classification as a conal component \citep{1999ApJS..121..171W}.  
We note that the decrease in the pulse
profile component widths with increasing frequency is
consistent with the RFM model
\citep[e.g.][]{2002ApJ...565..500G}, although this may also be caused by other mechanisms \citep[e.g. birefringence,][]{1997ApJ...475..763M}.  

It seems that the emission
behaviour observed from {\psr} is very diverse, and may prove one of
the most difficult pulsars to model in terms of accounting for the
host of emission characteristics that arise over many different
timescales. However, observations of {\psr} may also help in efforts to unify the diverse phenomenology of  pulsar emission
and provide vital data in terms of studies of the elusive pulsar emission mechanism.

\subsection{Pulsar Population Implication}

In the period--period-derivative ($P$--{$\dot P$}) diagram for non-recycled
pulsars from the ATNF pulsar catalogue \citep{2005AJ....129.1993M},
{\psr} is located near the centre of the distribution, and therefore
appears to be a very regular pulsar in this parameter space.  Other
pulsars which have been identified to display at least one of the
numerous emission characteristics investigated here are located
throughout most of the parameter space, apart from the area occupied
by very young pulsars. Therefore, there is no obvious correlation with
surface magnetic field strength or characteristic age
\citep[e.g.][]{2007A&A...469..607W}.  Most pulsars may exhibit these
emission phenomena, but the effects of pulse-to-pulse variability are
diminished due to summing over many pulses, which is required to achieve
sufficient S/N in the case of sources with low flux densities. 

To date, only a handful out of the $\sim 2300$ known pulsars are
identified as intermittent or extreme nullers.  This is because this
requires frequent observations over several months/years to identify
this long-term behaviour.  This is also because they may also be
difficult to identify in a pulsar survey due to the nature of the
weak/undetectable emission during quiet modes.  
There may also be pulsars that emit in a Q-mode-like state
over longer timescales or continuously, for which more sensitive
radio-telescopes will be needed, such as the Square Kilometre Array
\citep[SKA,][]{2010iska.meetE..18G}.  Studying pulsars that display
many of these emission characteristics, such as {\psr}, will provide
further insights into the possible radio emission characteristics, the
relationship between phenomena, and understanding of the
magnetospheric emission mechanism from pulsars and RRATs.

\subsection{Using Pulsars as ISM Probes}\label{subsec:probe}

Using the LOFAR observation on 2011 November 14, we found that there is no significant variation in the observed RM
over small timescales due to changes in the pulsar magnetosphere.
This is reassuring in terms of using pulsar RMs as probes of the
Galactic magnetic field \citep[e.g.][]{2011ApJ...728...97V}.  Although the pulsar magnetosphere is
assumed to have a minor contribution to the observed RM
\citep{2009MNRAS.396.1559N}, an interesting area of future
investigation may be whether the RM, or polarisation angle, as a
function of pulse phase is in any way dependent on the emission mode.
LOFAR's low frequency and large fractional bandwidth can facilitate such
studies, where high RM precision is important.  Studies of the emission
characteristics from pulsars, including in polarisation, will also
continue to aid in better understanding the pulsar emission mechanism
itself, so that pulsars can be used increasingly effectively as probes
of the ISM.

\section*{Acknowledgments}

We thank the Deutsche Forschungsgemeinschaft (DFG) for funding this
work within the research unit FOR 1254 `Magnetisation of Interstellar
and Intergalactic Media: The Prospects of Low-Frequency Radio
Observations'.  N.Y. acknowledges support from the National Research
Foundation (NRF).  J.W.T.H. acknowledges funding from an NWO Vidi
fellowship and ERC Starting Grant `DRAGNET' (337062).
The Low-Frequency Array (LOFAR) was designed and
constructed by ASTRON, the Netherlands Institute for Radio Astronomy,
and has facilities in several countries that are owned by various
parties (each with their own funding sources), and that are
collectively operated by the International LOFAR Telescope (ILT)
foundation under a joint scientific policy. The WSRT is operated by
ASTRON/NWO.  Observations with the Lovell Telescope are supported through an STFC consolidated grant.
This work is partly based on observations with the 100-m
telescope of the Max-Planck-Institut f\"{u}r Radioastronomie
at Effelsberg. We would like to thank R. Karuppusamy for his help with the Effelsberg observations.
C.F. acknowledges financial support by the {\it ``Agence Nationale de la Recherche''} through grant ANR-09-JCJC-0001-01.
The majority of the plots were created using the \textsc{PSRCHIVE} \textsc{Python} interface, and the
\textsc{Python} package \textsc{matplotlib} \citep{Hunter:2007}.

\bibliographystyle{mn2e} 
\bibliography{B0823+26_mnras_v8}

\label{lastpage}

\end{document}